\documentclass[pdflatex]{llncs}
\usepackage{etex}
\usepackage{hyperref} 
\usepackage{epsfig}
\usepackage{amssymb,stmaryrd,proof,subfigure,amsmath}
\usepackage[all]{xy}
\usepackage{changebar}
\newcommand{\dash}{\nobreakdash-}

\newif\iffull
 \fulltrue

\iffull
\usepackage{a4wide}
\newcommand{\short}[1]{}
\newcommand{\full}[1]{#1}
\else
\newcommand{\short}[1]{#1}
\newcommand{\full}[1]{}
\fi
   
\usepackage{ifthen}

\newcounter{ncomm}

\newcommand{\sem}[2][]{\ensuremath{\{\!| {#2} |\!\}^{#1}}}

\newcommand{\lc}{\mathcal{L}}
\newcommand{\lcwf}{\mathcal{L}_{\mathit{wf}}}

\newcommand{\lchm}{\mathcal{L}_{HM}}
\newcommand{\lcs}{\mathcal{L}_s}
\newcommand{\lcp}{\mathcal{L}_p}
\newcommand{\lchp}{\mathcal{L}_{hp}}

\newcommand{\lcmu}{\mu\mathcal{L}}

\newcommand{\stepb}{\ensuremath{\sim_\mathit{s}}}
\newcommand{\pomb}{\ensuremath{\sim_\mathit{p}}}
\newcommand{\hpb}{\ensuremath{\sim_\mathit{hp}}}
\newcommand{\hhpb}{\ensuremath{\sim_\mathit{hhp}}}

\newcommand{\pomsets}[1]{\ensuremath{Pom}(#1)}
\newcommand{\pom}{p}

\newcommand{\act}[1][a]{\mathsf{#1}}

\newcommand{\en}[4][\mathsf{a}]{
  \ifthenelse{\equal{#2}{}}
  {\ifthenelse{\equal{#3}{}}
    {\ensuremath{(\act[{#1}\,]{#4})}}
    {\ensuremath{(\overline{#3} <\act[{#1}\,]{#4})}}}
  {\ifthenelse{\equal{#3}{}}
    {\ensuremath{({#2}<\act[{#1}\,]{#4})}}
    {\ensuremath{({#2}, \overline{#3} <\act[{#1}\,]{#4})}}
  }}
\newcommand{\enBox}[4][\mathsf{a}]{
  \ifthenelse{\equal{#2}{}}
  {\ifthenelse{\equal{#3}{}}
    {\ensuremath{\{\act[{#1}\,]{#4}\}}}
    {\ensuremath{\{\overline{#3} <\act[{#1}\,]{#4}\}}}}
  {\ifthenelse{\equal{#3}{}}
    {\ensuremath{\{{#2}<\act[{#1}\,]{#4}\}}}
    {\ensuremath{\{{#2}, \overline{#3} <\act[{#1}\,]{#4}\}}}
  }}

\newcommand{\enx}[4][\mathsf{a}]{ 
  \ifthenelse{\equal{#2}{}}
  {\ifthenelse{\equal{#3}{}}
    {\ensuremath{\langle\!|\act[{#1}\,]{#4}|\!\rangle}}
    {\ensuremath{\langle\!|\overline{#3} < \act[{#1}\,]{#4}|\!\rangle}}}
  {\ifthenelse{\equal{#3}{}}
    {\ensuremath{\langle\!|{#2}<\act[{#1}\,]{#4}|\!\rangle}}
    {\ensuremath{\langle\!|{#2}, \overline{#3} <\act[{#1}\,]{#4}|\!\rangle}}
  }}

\newcommand{\enxBox}[4][\mathsf{a}]{ 
  \ifthenelse{\equal{#2}{}}
  {\ifthenelse{\equal{#3}{}}
    {\ensuremath{[\![\act[{#1}\,]{#4}]\!]}}
    {\ensuremath{[\![\overline{#3} < \act[{#1}\,]{#4}]\!]}}}
  {\ifthenelse{\equal{#3}{}}
    {\ensuremath{[\![{#2}<\act[{#1}\,]{#4}]\!]}}
    {\ensuremath{[\![{#2}, \overline{#3} <\act[{#1}\,]{#4}]\!]}}
  }}

\newcommand{\ex}[1]{\ensuremath{\langle {#1}\rangle}\,}

\newcommand{\exBox}[1]{\ensuremath{[\, {#1}\,]}\,}

\newcommand{\true}{\ensuremath{\mathsf{T}}}
\newcommand{\false}{\ensuremath{\mathsf{F}}}

\newcommand{\var}[1]{\ensuremath{\vec{#1}}}
\newcommand{\vx}{{\var{x}}}
\newcommand{\vy}{{\var{y}}}

\newcommand{\aProp}{\mathcal{X}^a}
\newcommand{\Prop}{\mathcal{X}}
\newcommand{\p}[2]{\ensuremath{{#1}({#2})}}

\newcommand{\Env}{\ensuremath{\mathit{Env}}}
\newcommand{\PEnv}{\ensuremath{\mathit{PEnv}}}

\newcommand{\fv}[1]{\ensuremath{\mathit{fv}({#1})}}
\newcommand{\fp}[1]{\ensuremath{\mathit{fp}({#1})}}

\newcommand{\art}[1]{\ensuremath{\mathit{ar}({#1})}}

\newcommand{\ld}[2][]{\ensuremath{\mathit{lp}_{#1}({#2})}}

\newcommand{\Var}{\ensuremath{\mathit{Var}}}

\newcommand{\trans}[1]
{ \setbox0=\hbox{$\ {}^{#1}\ $}
 \setbox1=\hbox{$\longrightarrow$}
 \loop\setbox1=\hbox{$-$\kern-0.3em\unhbox1}\ifdim\wd1<\wd0\repeat
 \hbox{$\ \ \mathop{\box1}\limits^{#1}\ \ $}
}

\newcommand{\ntrans}[1]
{ \setbox0=\hbox{$\ {}^{#1}\ $}
 \setbox1=\hbox{$\longrightarrow_n$}
 \loop\setbox1=\hbox{$-$\kern-0.3em\unhbox1}\ifdim\wd1<\wd0\repeat
 \hbox{$\ \ \mathop{\box1}\limits^{#1}\ \ $}
}

\newcommand{\Ntrans}[1]
{ \setbox0=\hbox{$\ {}^{#1}\ $}
 \setbox1=\hbox{$\not\longrightarrow$}
 \loop\setbox1=\hbox{$-$\kern-0.3em\unhbox1}\ifdim\wd1<\wd0\repeat
 \hbox{$\ \ \mathop{\box1}\limits^{#1}\ \ $}
}

\newcommand{\leftQ}{\{\kern-.22em|\,}
\newcommand{\rightQ}{\,|\kern-.22em\}}

\newcommand{\pes}{\textsc{pes}}

\newcommand{\conf}[1]{\ensuremath{\mathcal{C}({#1})}}

\newcommand{\res}[2]{\ensuremath{{#1}[{#2}]}}

\newcommand{\conc}{\ensuremath{\mathop{||}}}

\newcommand{\cons}{\ensuremath{\mathop{\smallfrown}}}

\newcommand{\cause}[1]{\ensuremath{\lceil {#1} \rceil }}

\newcommand{\sep}{\ensuremath{\mathop{\otimes}\,}}

\title{A Logic for True Concurrency\thanks{Work partially supported by the MIUR-PRIN Project CINA.}}
\author{Paolo Baldan and Silvia Crafa}

\institute{Department of Mathematics, University of Padova}

\begin{document}

\maketitle

\begin{abstract}
  We propose a logic for true concurrency whose formulae predicate
  about events in computations and their causal dependencies.
  %
  %
  %
  The induced logical equivalence is hereditary history preserving
  bisimilarity, and fragments of the logic can be identified which
  correspond to other true concurrent behavioural equivalences in the
  literature: step, pomset and history preserving
  bisimilarity. Standard Hennessy-Milner logic, and thus (interleaving)
  bisimilarity, is also recovered as a fragment.
  We also propose an extension of the logic with fixpoint operators,
  thus allowing to describe causal and concurrency properties of
  infinite computations.
  %
  %
  This work contributes to a rational presentation of the true
  concurrent spectrum and to a deeper understanding of the relations
  between the involved behavioural equivalences.
\end{abstract}

\section{Introduction}

In the semantics of concurrent and distributed systems, a major
dichotomy opposes the interleaving approaches, where concurrency of
actions is reduced to the non-deterministic choice among their
possible sequentialisations, to true concurrent approaches, where
concurrency is taken as a primitive notion.
In both cases, on top of the operational models a number of
behavioural equivalences have been defined by abstracting from aspects which
are considered unobservable~\cite{vGlabSequential,vGlab01}.

For the interleaving world, a systematic and impressive picture is
taken in the linear-time branching-time
spectrum~\cite{vGlabSequential}.
Quite interestingly, the equivalences in the spectrum can be uniformly
characterised in logical terms.  Bisimilarity, the finest equivalence,
corresponds to Hennessy-Milner (HM) logic: two processes are bisimilar
if and only if they satisfy the same HM logic
formulae~\cite{HM:ALCND}.
Coarser equivalences correspond to suitable fragments of HM logic, as
discussed in~\cite{vGlabSequential}.

In the true concurrent world, relying on models like event structures
or transition systems with independence~\cite{NielWin:Models}, several
behavioural equivalences have been defined.
Hereditary history preserving (hhp-)bisimilarity~\cite{Ber:HHP}, the finest
equivalence in the spectrum in~\cite{vGlab01}, has been shown to
arise as a canonical behavioural equivalence when considering
partially ordered computations~\cite{JNW:BFOM} (Their abstract notion of bisimilarity instantiates to hhp-bisimilarity when taking the category of pomsets as the path category.)
Coarser equivalences like history preserving (hp\dash)bisimilarity~\cite{RT:BSN,DDM:POD,BDKP:FCB}, pomset and
step bisimilarity have also been widely studied.
Correspondingly, a number of logics have been
studied, but, to the best of our knowledge, a unifying logical
framework for the main true concurrent equivalences is still
missing.
The huge amount of work on the topic makes it impossible to give a
complete account of related approaches.
Just to give a few references (see Section~\ref{se:conclude} for a
wider discussion),~\cite{DF:OLCM} proposes a general framework
encompassing a number of temporal and modal logics that characterise
interleaving bisimilarity as well as pomset bisimilarity and weak
hhp-bisimilarity, a weakening of hhp-bisimilarity studied, e.g., in~\cite{DF:OLCM,LPS:TechRep94,Cher:BFbis}.
However, finer equivalences are not
considered and a single unitary logic is missing. 
Hp-bisimilarity has been studied in the
setting of Petri nets and shown to be decidable for finite 1-safe
Petri nets in~\cite{Vog:DHPB}. A decidability result for finite-state
Petri nets is obtained also in~\cite{MP:MTS} by means of an encoding
of into history dependent (HD\dash)automata.
Concerning hhp-bisimilarity, several logics with modalities
corresponding to the ``retraction'' or ``backward'' execution of
computations have been proposed~\cite{HS:PFP,Ber:HHP,NC:GLNB,PU:LRM}. When a system does not exhibit autoconcurrency, i.e., when two instances of the same action are never enabled in parallel,
such logics are shown to capture hhp-bisimilarity. Relaxing this restriction  requires to move to an event based logic, where specific events executed in the past can be retracted~\cite{Ber:HHP,NC:GLNB,PU:LRM}.

In this paper we propose a behavioural logic for concurrency and we
show that it allows us to characterise a relevant part of the true
concurrent spectrum.
More specifically, the full logic $\lc$ is shown to capture
hhp-bisimilarity, the finest behavioural equivalence in the spectrum
in~\cite{vGlab01}.
Then suitable fragments of the logic are shown to scale down to the
characterisation of other coarser equivalences: history preserving,
pomset and step bisimilarity.
Standard HM logic, and thus (interleaving) bisimilarity, is also
recovered as a fragment.

Our logic allows us to predicate about events in computations together
with their causal and independence relations. It is interpreted over
prime event structures~\cite{NPW:PNES,Win:ES}, one of the most widely
known event-based models of computation, where the dependencies
between events are expressed in terms of causality and (binary)
conflict. It could naturally be interpreted over any formalism with
explicit notions of event, causality and consistency.
A formula is evaluated in a configuration representing the current
state of the computation, and it predicates on the possible future
evolutions starting from that state.
The logic is event-based in the sense that it contains an operator
acting as a binder: it asserts the existence of an event satisfying
suitable requirements and it binds the event to a variable so that the
event can be referred to later in the formula. In this respect, it is
reminiscent of the modal analogue of independence-friendly modal logic
as considered in~\cite{BF:IndepLogic02}.  

The logic contains two main operators. The formula
$\en{x}{y}{z}\varphi$ declares that an $\act$-labelled future event
exists, which causally depends on the event bound to $x$, and is
independent from the event bound to $y$.
Such an event is bound to variable $z$ so that it can be later
referred to in $\varphi$. In general, $x$ and $y$ can be replaced by
tuples of variables.
A second operator allows one to ``execute'' events previously bound to
variables.  The formula $\ex{z}\varphi$ says that the event bound to
$z$ is enabled in the current state, and after its execution $\varphi$
holds.

Different behavioural equivalences are induced by fragments of the
logics where we suitably restrict the set of possible futures the
formulae are able to refer to.
Namely, hhp-bisimilarity, that is captured by the full logic,
corresponds to the ability of observing the existence of a number of
legal but (possibly) incompatible futures. Such ability is strictly
related to the capability of observing future events without executing
them (in fact the execution of an event would rule out all the events
in conflict with it).
Interestingly, the definition of hhp-bisimilarity is normally given in
terms of backward transitions, whereas our logical
characterisation has a ``forward flavour.'' 
By restricting to a
fragment where future events can be observed only 
by executing them (any occurrence of the binding operator is 
immediately followed by a corresponding execution), we get
hp-bisimilarity.
Pomset bisimilarity is induced by a fragment of the logic obtained by
further restricting that for hp-bisimilarity, with the requirement
that propositional connectives are used only on closed
(sub)formulae. Roughly speaking, this fragment predicates about the
possibility of executing pomset transitions and the closedness
requirement prevents pomset transitions from being causally linked to
the events in the past.
Finally, step bisimilarity corresponds to the
possibility of observing only currently enabled concurrent events.
%

%

The logic $\lc$ in its basic form is essentially a means to understand
and compare different process equivalences, but its expressive
power is rather weak.  In fact, although events arbitrarily far in the
future can be ``observed'', the logic only allows us to describe
computations where a finite number of events are executed.
In order to overcome this limitation and to provide a more powerful
specification logic, well-suited for describing properties of
unbounded, possibly infinite computations, we enrich the logic with a
form of recursion. This is obtained by adding least (and dually
greatest) fixpoint operators, thus obtaining a kind of first order
modal $\mu$-calculus similar to the $\mu$-calculi
in~\cite{Dam:MCMP,DFG:TPVODS} and~\cite{GW:MCPD}, which are endowed
with first order variables representing channels or data. Similarities
exist also with the fixpoint extension of independence-friendly modal
logic in~\cite{BK:CIFFL}.
In the resulting logic $\lcmu$ one can express non-trivial causal
properties, like ``any $\act$ action can always be followed by a
causally related $\act[b]$ action in at most three steps,'' or ``an
$\act$ action can always be executed in parallel with a $\act[b]$
action.'' Moreover, we show that, as it happens in the interleaving
case, the addition of the fixpoint operators does not alter the
logical equivalence. The logical equivalence of $\lcmu$ is still
hhp-bisimilarity and the same invariance result applies to the
fixpoint extensions of the fragments of $\lc$ characterising the
coarser behavioural equivalences.

This work contributes to the definition of a logical counterpart of
the true concurrent spectrum, shading further light on the relations
between the involved behavioural equivalences and suggests interesting
directions of investigations in the verification of true concurrent
properties.

The rest of the paper is organised as follows. In
Section~\ref{se:back} we introduce the basics of event structures and
the concurrent equivalences we will work with in the paper. In
Section~\ref{se:logic} we present the syntax and semantics of our
logic $\lc$. In Section~\ref{se:hhp-characterisation} we study the
logical equivalence induced by $\lc$, proving that it
coincides with hhp-bisimilarity. In Section~\ref{se:spectrum} we
provide a characterisation of other concurrent equivalences in terms
of fragments of our logic.
In Section~\ref{se:recursion} we discuss the fixpoint extension of our logic. 
Finally, in Section~\ref{se:conclude} we discuss some
related work and present directions of future research.
This is a revised and extended version of the conference paper~\cite{BC:LTC}.

\section{Background}
\label{se:back}

In this section we provide the basics of prime event structures which
will be used as models for our logic. Then we define some common
behavioural true concurrent equivalences which will play a basic role
in the paper.

\subsection{Event structures}

Prime event structures~\cite{NPW:PNES,Win:ES} are a widely known model of
concurrency. They describe the behaviour of a system in terms of
events and dependency relations between such events. Throughout the
paper $\Lambda$ denotes a fixed set of labels ranged over by
$\act, \act[b], \act[c]$ \ldots

\begin{definition}[prime event structure]
  A ($\Lambda$-labelled) \emph{prime event structure} ({\pes}) is a
  tuple $\mathcal{E} = \langle E, \leq, \#, \lambda \rangle$, where
  $E$ is a denumerable set of \emph{events}, $\lambda : E \to \Lambda$ is a
  labelling function and $\leq$, $\#$ are binary relations
  on $E$, called \emph{causality} and \emph{conflict} respectively,
  such that:
  \begin{enumerate}
    \item  $\leq$ is a partial order and $\cause{e}
      = \{ e' \in E \mid e' \leq e \}$ is finite for all $e \in E$;
    \item $\#$ is irreflexive, symmetric and hereditary
      with respect to $\leq$, i.e., for all $e,e',e'' \in E$, if $e \#
      e' \leq e''$ then $e \# e''$.
  \end{enumerate}
\end{definition}

In the following, we will assume that the components of an event
structure $\mathcal{E}$ are named as in the definition
above. Subscripts carry over the components.

\begin{definition}[consistency, concurrency]
  Let $\mathcal{E}$ be a {\pes}.  We say that $e, e' \in E$ are
  \emph{consistent}, written $e \cons e'$, if $\neg (e \# e')$.
  A subset $X \subseteq E$ is called \emph{consistent} if $e \cons
  e'$ for all $e, e' \in X$. 
  %
  %
  We say that $e$ and $e'$ are \emph{concurrent}, written $e \conc
  e'$, if $\neg (e \leq e')$, $\neg (e' \leq e)$ and $\neg (e \# e')$.
\end{definition}

Causality, concurrency and consistency will be sometimes used on sets of
events. Given $X \subseteq E$ and $e \in E$, by $X < e$ we mean that
for all $e' \in X$, $e' < e$. Similarly $X \conc e$, resp. $X \cons
e$, means that for all $e' \in X$, $e' \conc e$, resp. $e' \cons e$.
We write $\cause{X}$ for $\bigcup_{e \in X} \cause{e}$.

Configurations of event structures are intended to represent
(concurrent) computations, which abstract from the order of execution
of concurrent events.

\begin{definition}[configuration]
  Let $\mathcal{E}$ be a {\pes}. 
  A \emph{(finite) configuration} in $\mathcal{E}$ is a (finite)
  consistent subset of events $C\subseteq E$ closed
  w.r.t. causality (i.e., $\cause{C} = C$).
  The set of finite configurations of $\mathcal{E}$ is denoted by
  $\conf{\mathcal{E}}$.
\end{definition}

Observe that the empty set of events $\emptyset$ is always a
configuration, which can be understood as the initial state of the
computation.
  
Hereafter all configurations will be assumed to be finite. A
consistent subset $X \subseteq E$ of events will always be seen as a
\emph{pomset} (partially ordered multiset) $(X, \leq_X, \lambda_X)$,
where $\leq_X$ and $\lambda_X$ are the restrictions of $\leq$ and
$\lambda$ to $X$. Given $X, Y \subseteq E$ we will write $X \sim Y$ if
$X$ and $Y$ are isomorphic as pomsets.

\begin{definition}[pomset transition and step]
  Let $\mathcal{E}$ be a {\pes} and let $C \in
  \conf{\mathcal{E}}$. 
  Given $\emptyset \neq X \subseteq E$, if $C \cap X = \emptyset$ and
  $C' = C \cup X \in \conf{\mathcal{E}}$ we write $C \trans{X} C'$ and
  call it a \emph{pomset transition} from $C$ to $C'$. When the events
  in $X$ are pairwise concurrent,
  we say that $C \trans{X} C'$ is a \emph{step}. When $X = \{ e \}$ we
  write $C \trans{e} C'$ instead of $C \trans{\{e\}} C'$.
\end{definition}

A {\pes} $\mathcal{E}$ is called \emph{image finite} if for any $C \in
\conf{\mathcal{E}}$ and $\act \in \Lambda$, the set of events $\{ e \in
E \mid C \trans{e} C'\ \land\ \lambda(e) = \act \}$ is finite.
\emph{All} the {\pes}s considered in this paper will be assumed to be
image finite. As it commonly happens when relating modal logics and
bisimilarities, this assumption is crucial for getting a logical
characterisation of the various bisimulation equivalences in
Sections~\ref{se:hhp-characterisation} and~\ref{se:spectrum}, based on
a finitary logic.

\subsection{Concurrent   behavioural equivalences}

Behavioural equivalences which capture to some extent the concurrency
features of a system, can be defined on the transition system where
states are configurations and transitions are pomset transitions. 

\begin{definition}[pomset, step bisimulation]
  Let $\mathcal{E}_1$, $\mathcal{E}_2$ be {\pes}s. A \emph{pomset
    bisimulation} is a relation $R \subseteq \conf{\mathcal{E}_1}
  \times \conf{\mathcal{E}_2}$ such that if $(C_1, C_2) \in R$ and
  $C_1 \trans{X_1} C_1'$ then $C_2 \trans{X_2} C_2'$, with $X_1 \sim
  X_2$ and $(C_1', C_2') \in R$, and vice versa.  We say that
  $\mathcal{E}_1$, $\mathcal{E}_2$ are \emph{pomset bisimilar},
  written $\mathcal{E}_1 \pomb \mathcal{E}_2$, if there exists a
  pomset bisimulation $R$ such that $(\emptyset, \emptyset) \in R$.

  \emph{Step bisimulation} is defined analogously, replacing general
  pomset transitions with steps. We write $\mathcal{E}_1 \stepb
  \mathcal{E}_2$ when $\mathcal{E}_1$ and $\mathcal{E}_2$ are step
  bisimilar.
\end{definition}

While pomset and step bisimilarity only consider the causal structure
of the current step, (hereditary) history preserving bisimilarities
are sensible to the way in which the executed events depend on events
in the past. In order to define history preserving bisimilarities the
following definition is helpful.

\begin{definition}[posetal product]
  Given two {\pes}s $\mathcal{E}_1$, $\mathcal{E}_2$, the
  \emph{posetal product} of their configurations, denoted
  $\conf{\mathcal{E}_1} \bar{\times} \conf{\mathcal{E}_2}$, is defined
  as
  \begin{center}
    $\{ (C_1, f, C_2) \mid C_1 \in \conf{\mathcal{E}_1},\ 
    C_2 \in \conf{\mathcal{E}_2},\ 
    f : C_1 \to C_2\ \mathrm{isomorphism} \}$ 
  \end{center}
  A subset $R \subseteq \conf{\mathcal{E}_1} \bar{\times}
  \conf{\mathcal{E}_2}$ is called a \emph{posetal relation}. We
  say that $R$ is downward closed when for any $(C_1, f, C_2), (C_1',
  f', C_2') \in \conf{\mathcal{E}_1} \bar{\times}
  \conf{\mathcal{E}_2}$, if $(C_1, f, C_2) \subseteq (C_1', f',
  C_2')$ pointwise and $ (C_1', f', C_2') \in R$ then $ (C_1, f, C_2)
  \in R$.
\end{definition}

Given a function $f: X_1 \to X_2$ we will denote by $f[x_1 \mapsto
  x_2] : X_1 \cup \{ x_1 \} \to X_2 \cup \{ x_2 \}$ the function
defined, for $z \in X_1 \cup \{x_1\}$, by
\begin{center}
  $f[x_1 \mapsto x_2] (z) = \left\{
  \begin{array}{ll}
    x_2 & \mbox{if $z = x_1$}\\[1mm]
    f(z) & \mbox{otherwise}
  \end{array}
  \right.
  $
\end{center}

\begin{definition}[(hereditary) history preserving bisimulation]
  A \emph{history preserving (hp-)bisimulation} is a posetal relation
  $R \subseteq \conf{\mathcal{E}_1} \bar{\times} \conf{\mathcal{E}_2}$
  such that if $(C_1, f, C_2) \in R$ and $C \trans{e_1} C_1'$ then
  $C_2 \trans{e_2} C_2'$, with $(C_1', f[e_1 \mapsto e_2], C_2') \in
  R$, and vice versa.
  We say that $\mathcal{E}_1$, $\mathcal{E}_2$ are \emph{history
    preserving (hp-)bisimilar} and write $\mathcal{E}_1 \hpb
  \mathcal{E}_2$ if there exists a hp-bisimulation $R$ such that
  $(\emptyset, \emptyset, \emptyset) \in R$.

  A \emph{hereditary history preserving (hhp-)bisimulation} is a
  downward closed hp-bisimulation.
  The fact that $\mathcal{E}_1$,
  $\mathcal{E}_2$ are \emph{hereditary history preserving
    (hhp-)bisimilar} is denoted $\mathcal{E}_1 \hhpb \mathcal{E}_2$.
\end{definition}

It is easy to see  (\cite{vGlab01}) that
the definition of (h)hp-bisimilarity can be
equivalently given by using pomset transitions instead of single event
transitions, i.e., by asking that if $(C_1, f, C_2) \in R$ and $C
\trans{X_1} C_1'$ then there exists $C_2 \trans{X_2} C_2'$ and
$(C_1', f', C_2') \in R$, with $f'_{|C_1} = f$.

\section{A logic for true concurrency}
\label{se:logic}

In this section we introduce the syntax and the semantics of our
logic. Formulae predicate about events in computations and their
dependencies as primitive concepts.
The logic is interpreted over {\pes}s. It could be interpreted,
without any serious technical complication, over more general classes
of event structures, as long as they are endowed with notions of
causality and consistency (e.g., over stable event
structures~\cite{Win:ES}). The choice of restricting to {\pes} is
motivated by the fact that they are probably the most popular event
structure model, easily accessible and, at the same time, quite
expressive.

In order to keep the notation simple, tuples of variables like $x_1,
\ldots, x_n$ will be denoted by $\vx$
and, abusing the notation, tuples will be often used as sets.

\begin{definition}[syntax]
  \label{de:syntax}
  Let $\Var$ be a denumerable set of variables ranged over by $x,y,z,
  \ldots$. The syntax of the logic $\lc$ over the set of labels
  $\Lambda$ is defined as follows, where $\act$ ranges over $\Lambda$:
  \begin{center}
    $
    \varphi \ ::= \  
    \true \ \mid\ \varphi \land \varphi\ \mid\ \neg \varphi\ \mid\ \en{\vx}{\vy}{z}\, \varphi\ \mid\ \ex{z}\varphi
      %
    $
  \end{center}
\end{definition}
The operator $\en{\vx}{\vy}{z}$ 
acts as a binder for the variable $z$, as
clarified by the following notion of free variables in a formula.

\begin{definition}[free variables]
  \label{de:free-vars}
  The set of free variables of a formula $\varphi$, denoted
  $\fv{\varphi}$, is inductively defined by:
  $$
 \begin{array}{lll}
    \fv{\true}  & = & \emptyset\\[1mm]
    %
    \fv{\varphi_1\land\varphi_2} & = &  \fv{\varphi_1} \cup \fv{\varphi_2}\\[1mm]
    %
    %
    \fv{\neg \varphi} & = &  \fv{\varphi}\\[1mm]
    %
    %
    \fv{\en{\vx}{\vy}{z}\, \varphi} & = &
    \vx \cup \vy \cup (\fv{\varphi} \setminus \{ z \})\\[1mm]
    %
    %
    \fv{\ex{z}\varphi} & =  & \fv{\varphi}\cup\{z\}\\[1mm]
    %
  \end{array}
  $$
\end{definition}

The satisfaction of a formula $\varphi$ is defined with respect to a
configuration $C \in \conf{\mathcal{E}}$, representing the state of the
computation, and a (total) function $\eta : \Var \to E$, called an
\emph{environment}, that binds free variables in $\varphi$ to events
in $C$ or in the future of $C$. In particular, the events bound to
free variables in a formula must be both pairwise consistent and
consistent with the current state of the computation. Such a
requirement is expressed by the following definition of legal pair.

\begin{definition}[environments, legal pairs]
  \label{de:legal-pairs}
  Let $\mathcal{E}$ be a {\pes}.  We denote by $\Env_\mathcal{E}$ the
  set of environments $\eta : \Var \to E$. 
  Given a formula $\varphi$ in $\lc$, a pair $(C, \eta) \in
  \conf{\mathcal{E}} \times \Env_\mathcal{E}$ is \emph{legal} for
  $\varphi$ if $C \cup \eta(\fv{\varphi})$ is a consistent set of
  events.  We denote by $\ld[\mathcal{E}]{\varphi}$ the set of legal
  pairs for $\varphi$ in $\mathcal{E}$.
\end{definition}

\paragraph{Remark.}{ 
  Observe that the legal pairs for a formula only
  depends on its set of free variables. Whenever $\fv{\varphi} =
  \fv{\psi}$ it holds that $\ld[\mathcal{E}]{\varphi} =
  \ld[\mathcal{E}]{\psi}$. More generally, if $\fv{\varphi} \subseteq
  \fv{\psi}$ then $\ld[\mathcal{E}]{\varphi} \supseteq
  \ld[\mathcal{E}]{\psi}$.
}

We simply write $\Env$ and $\ld{\varphi}$, omitting the
subscript, when the {\pes} $\mathcal{E}$ is clear from the context.
%
%
Moreover, in order to simplify the definition of the semantics, 
given a configuration $C$, we denote by $\res{E}{C}$ the
\emph{residual} of $E$ after $C$, defined as 
$\res{E}{C}=\{ e \mid e \in E \setminus C\ \land\ C \cons e \}$.

\begin{definition}[semantics]
  \label{de:semantics}
  Let $\mathcal{E}$ be a {\pes}. The denotation of a formula
  $\varphi$, written $\sem[\mathcal{E}]{\varphi} \in
  2^{\conf{\mathcal{E}} \times \Env_{\mathcal{E}}}$ is defined inductively as follow:
  \begin{center}
    $
    \begin{array}{rcl}
      \sem[\mathcal{E}]{\true} & = &  \conf{\mathcal{E}} \times
      \Env_{\mathcal{E}}
      \\[3mm]
      %
      %
      \sem[\mathcal{E}]{\varphi_1 \land \varphi_2} & = &  
      \sem[\mathcal{E}]{\varphi_1} \cap \sem[\mathcal{E}]{\varphi_2} \cap \ld{\varphi \land \psi}\\[3mm]
      \sem[\mathcal{E}]{\neg \varphi} & = &  \ld{\varphi} \setminus \sem[\mathcal{E}]{\varphi}\\[3mm]
      \sem[\mathcal{E}]{\en{\vx}{\vy}{z}\, \varphi} & = &
      \{ (C, \eta) \mid 
      \begin{array}[t]{ll}
           (C, \eta) \in \ld{\en{\vx}{\vy}{z}\, \varphi}\ \mathrm{and}\\
          \exists e \in \res{E}{C} \mathrm{\ such\ that}\ e \cons \eta(\fv{\varphi} \setminus \{ z \})\\
          \land\ \lambda(e) = \act\ \land\ \eta(\vx) < e\ \land\ \eta(\vy) \conc e\\
           \land\  (C, \eta[z\mapsto e]) \in
           \sem[\mathcal{E}]{\varphi}  & \}\\[3mm]
      \end{array}
      \\
      %
      %
      \sem[\mathcal{E}]{\ex{z}\, \varphi} & = &
      \{ (C, \eta) \mid 
            \begin{array}[t]{ll}
          C \trans{\eta(z)} C'\ \land\  (C', \eta) \in
      \sem[\mathcal{E}]{\varphi} & \}
      \end{array}

      \\[2mm]
      %
      %
    \end{array}
    $
  \end{center}

  \noindent
  %
  When $(C, \eta) \in \sem[\mathcal{E}]{\varphi}$ we say that the
  {\pes} $\mathcal{E}$ satisfies the formula $\varphi$ in the
  configuration $C$ and environment $\eta : \Var \to E$, and write
  $\mathcal{E}, C \models_\eta \varphi$. For closed formulae
  $\varphi$, we write $\mathcal{E}, C \models \varphi$, when
  $\mathcal{E}, C \models_\eta \varphi$ for some $\eta$ and
  $\mathcal{E} \models \varphi$, when $\mathcal{E}, \emptyset \models
  \varphi$.
\end{definition}

Intuitively, the formula
\begin{center}
  $\en{\vx}{\vy}{z}\, \varphi$
\end{center}
holds in $(C,\eta)$ when in the future of the configuration $C$ there
is an $\act$-labelled event $e$, consistent with the events bound to
free variables in $\varphi$, such that binding $e$ to variable $z$, the
formula $\varphi$ holds. Such an event is required to be caused (at
least) by the events already bound to variables in $\vx$, and to be
independent (at least) from those bound to variables in $\vy$.
We stress that the event $e$ might not be currently enabled; it is
only required to be consistent with the current configuration, meaning
that it could be enabled in the future of the current configuration.
The formula $\ex{z}\varphi$ says that the event bound to $z$ is
enabled by the current configuration, hence it can be executed
producing a new configuration which satisfies the formula $\varphi$.
To simplify the notation we write $\en{}{}{z}\, \varphi$ for
$\en{\ }{}{z}\, \varphi$.


\begin{figure}[t]
\begin{center}
\begin{tabular}{ccccc}
\mbox{
 \xymatrix@R=3mm@C=5mm{
      b  &  d   \\
      a\ar@{-}[u]\ar@{.}[r]  & c\ar@{-}[u] \\
    }}
&\hspace*{1cm}
\mbox{
 \xymatrix@R=3mm@C=2.5mm{
      b  \ar@{.}[rr] & &  d   \\
      & a\ar@{-}[ur]\ar@{-}[ul] &\\
    }}
&\hspace*{1cm}
\mbox{
 \xymatrix@R=3mm@C=3.5mm{
      \\
      a & b \ar@{.}[r] &  d\\
    }}
&\hspace*{1cm}
\mbox{
\xymatrix@R=3mm@C=5mm{
      c & c \\
      b \ar@{-}[u] & b \ar@{-}[u]\\
      a \ar@{-}[u] \ar@{.}[r] & a \ar@{-}[u]\\
    }}
&\hspace*{8mm}
\mbox{
 \xymatrix@R=3mm@C=3mm{
      c \\
      b \ar@{-}[u]\\
      a \ar@{-}[u]
    }}
\\ \\
$\mathcal{E}_1$
&\hspace*{1cm}
$\mathcal{E}_2$
&\hspace*{1cm}
$\mathcal{E}_3$
&\hspace*{1cm}
$\mathcal{E}_4$
&\hspace*{8mm}
$\mathcal{E}_5$
\end{tabular}
\end{center}
\caption{}
  \label{fi:example1}
\end{figure}

As an example, consider the {\pes} $\mathcal{E}_1$ in
Fig.~\ref{fi:example1}, corresponding to the CCS process $a.b+c.d$,
where dotted lines represent immediate conflict and the causal order
proceeds upwards along the straight lines. The empty configuration
satisfies the closed formula $\en[b]{}{}{x} \true$, i.e.,
$\mathcal{E}_1 \models \en[b]{}{}{x} \true$, even if the
$\act[b]$-labelled event is not immediately enabled. Also
$\mathcal{E}_1 \models
\en[b]{}{}{x} \true \wedge \en[d]{}{}{y} \true$, since there
are two possible (incompatible) computations that start from the
empty configuration and contain, respectively, a $\act[b]$-labelled and a
$\act[d]$-labelled event. On the other hand, if
$\varphi=\en{}{}{z}\ex{z}(\en[b]{}{}{x}\true  \land \en[d]{}{}{y}) \true$
then $\mathcal{E}_1 \not\models \varphi$ since after the execution of
the $\act$-labelled event, $\mathcal{E}_1$ reaches a
configuration that does not admit a future containing an event
labelled by $\act[d]$. As a further example, the formula $\varphi$
above is satisfied by the {\pes}s $\mathcal{E}_2$ and $\mathcal{E}_3$
in Fig.~\ref{fi:example1} corresponding respectively to the process
$a.(b+d)$ and $a \mid (b+d)$, whereas the formula
$\en{}{}{z} \ex{z} \en[b]{}{z}{x} \true$ is satisfied only by
$\mathcal{E}_3$.

\medskip

It is worth noticing that the semantics of the binding operator does
not prevent from choosing for $z$ an event $e$ that has been already
bound to a different variable, i.e., the environment function $\eta$
need not be injective. This is essential to avoid the direct
observation of conflicts, a capability which would make the logical
equivalence stronger than hhp-bisimilarity (and of any reasonable
behavioural equivalence). Consider for instance the {\pes}s associated
to the hhp-equivalent processes $a+a$ and $a$: in order to be also
logically equivalent, they both must satisfy the formula $\en{}{}{z}
\en{}{}{z'} \true$. Hence for the second {\pes}, both $z$ and $z'$
must be bound to the unique $\act$-labelled event.  On the other hand,
observe that both {\pes}s falsify the formula
$\en{}{}{z}\en{}{}{z'}\ex{z}\ex{z'} \true$. In fact, $z'$ must
be bound to an event consistent with that associated to $z$
(because $z$ occurs free in $\ex{z}\ex{z'} \true$). Hence $z$
and $z'$ will be bound to the same event, which cannot be executed
twice.

\subsection{About legal pairs and environments}

We remark that differently from other logics for event structures,
whose semantics is given only with respect to the set of
configurations, here legal pairs come into play in order to ensure
that the events bound to free variables in a formula be consistent
with the current state of the computation and pairwise consistent. 
The intuition is that, in a legal pair for a formula, the
configuration identifies the current state of the computation and the
environment should map variables free in the formula to events which
have already occurred or which can occur in a possible future of the
current state.

The use of legal pairs has some subtle effects on the semantics of the
propositional connectives. In particular, concerning negation, it is
immediate to see that a pair $(C, \eta)$ is legal for $\varphi$ if and
only if it is legal for $\neg \varphi$. Hence, when a denotation $(C,
\eta)$ is not legal for $\varphi$, we have that neither $\mathcal{E},
C \models_\eta \varphi$ nor $\mathcal{E}, C \models_\eta \neg
\varphi$. As a concrete example, take $\varphi = \ex{x} \ex{y} \true$.
Then in the {\pes} $\mathcal{E}_1$ of Fig.~\ref{fi:example1}, if
$\eta$ binds $x$ and $y$ to the conflicting events labelled $\act$ and
$\act[c]$, respectively, then $(\emptyset, \eta)$ is not legal for
$\varphi$ and we have $\mathcal{E}_1, \emptyset \not\models_\eta
\varphi$ and $\mathcal{E}_1, \emptyset \not\models_\eta \neg \varphi$.


For closed formulae, we have the following:

\begin{lemma}[negation]
  Let $\varphi$ be a closed formula in $\lc$, let $\mathcal{E}$ be a {\pes}
  and let $(C, \eta) \in \conf{\mathcal{E}} \times \Env_\mathcal{E}$.  Then $\mathcal{E}, C
  \models_\eta \varphi$ iff $\mathcal{E}, C \not\models_\eta \neg
  \varphi$.
\end{lemma}

\begin{proof}
  Immediately follows from the fact that for a closed formula any pair
  is legal.
   \qed
\end{proof}

Concerning conjunction, observe that it is not the case that
$\ld{\varphi \land \psi} = \ld{\varphi} \cap \ld{\psi}$. Therefore it
can happen that $\mathcal{E}, C \models_\eta \varphi$ and
$\mathcal{E}, C \models_\eta \psi$, but $\mathcal{E}, C
\not\models_\eta \varphi \land \psi$. As an example, consider again
the {\pes} $\mathcal{E}_1$ of Fig.~\ref{fi:example1}, and the formulae
$\varphi = \ex{x} \true$ and $\psi = \ex{y} \true$. If $\eta$ binds
$x$ and $y$ to the events labelled $\act$ and $\act[c]$, respectively,
then $(\emptyset, \eta) \in \ld{\varphi}$, $(\emptyset, \eta) \in
\ld{\varphi}$ and we have $\mathcal{E}_1, \emptyset \models_\eta
\varphi$ and $\mathcal{E}_1, \emptyset \models_\eta \psi$. However,
since the two events are in conflict, $(\emptyset, \eta) \not\in
\ld{\varphi \land \psi}$, and thus $\mathcal{E}_1, \emptyset
\not\models_\eta \varphi \land \psi$.

We next show that the denotation of a formula, given according to
Definition~\ref{de:semantics}, always consists of a set of legal pairs
for the formula.

\begin{lemma}[denotations consist of legal pairs]
  \label{le:legal}
  Let $\mathcal{E}$ be a {\pes}. Then for any formula $\varphi \in
  \lc$, it holds $\sem[\mathcal{E}]{\varphi} \subseteq
  \ld[\mathcal{E}]{\varphi}$
\end{lemma}

\begin{proof}
  The proof is by routine  induction on the structure of the formula
  $\varphi$. We only comment case $\varphi = \ex{z} \psi$. If $(C,
  \eta) \in \sem[\mathcal{E}]{\varphi}$ then, by definition, if we let
  $e = \eta(z)$, it holds that $C \trans{e} C \cup \{ e \}$ and $(C
  \cup \{ e \}, \eta) \in \sem[\mathcal{E}]{\psi}$. Hence by inductive
  hypothesis $(C \cup \{ e \}, \eta) \in \ld[\mathcal{E}]{\psi}$,
  i.e., $C \cup \{e \} \cup \eta(\fv{\psi})$ is
  consistent. Since $\fv{\varphi} = \fv{\psi} \cup \{z\}$, we have
  that $C \cup \eta(\fv{\varphi}) = C \cup \{ e\} \cup
  \eta(\fv{\psi})$, and thus we can conclude $(C, \eta) \in
  \ld[\mathcal{E}]{\varphi}$.
  \qed
\end{proof}

The semantics of a formula only depends on the events that
the environment associates to the free variables of the formula.

\begin{lemma}
  \label{le:env-fv}
  Let $\mathcal{E}$ be a {\pes} and let $C \in \conf{E}$. Let $\varphi
  \in \lc$ and let $\eta_1, \eta_2 :
  \Var \to E$ be environments such that $\eta_1(x) = \eta_2(x)$ for
  any $x \in \fv{\varphi}$. Then 
  \begin{center}
    $\mathcal{E}, C \models_{\eta_1} \varphi$
    \qquad 
    iff
    \qquad
    $\mathcal{E}, C \models_{\eta_2} \varphi$
  \end{center}
  In particular,  $(C,\eta_1) \in \ld[\mathcal{E}]{\varphi}$ if
  and only if $(C,\eta_2) \in \ld[\mathcal{E}]{\varphi}$.
\end{lemma}

\begin{proof}
  Routine induction on the structure of $\varphi$.
  \qed
\end{proof}

Note that without restricting the semantics of formulae to legal pairs
the logics would have been too powerful. In fact, it would have
allowed us to observe conflicts through a combination of the binder
and the execution modality.
For instance, consider the {\pes}s $\mathcal{E}_4$ and $\mathcal{E}_5$
in Fig.~\ref{fi:example1}, corresponding to the processes
$a.b.c+a.b.c$ and $a.b.c$, respectively, and take formula $\varphi =
\en{}{}{x}\en[b]{}{}{y}\ex{x}\neg \ex{y} \true$, saying that there are
two events labelled by $\act$ and $\act[b]$ such that after executing
the first, the second cannot be executed. With the current definition
neither $\mathcal{E}_4$ nor $\mathcal{E}_5$ satisfy $\varphi$, since
after binding $x$ to any $\act$-labelled event $e$, in order to keep
the denotation legal, $y$ must be bound to the $\act[b]$-labelled
event caused by $e$, that is executable after $e$.
Without the restriction to legal pairs, instead, the formula
would hold in $\mathcal{E}_4$, since variables $x$ and $y$ could be
bound to conflicting events (e.g., $x$ could be bound to the
$\act$-labelled event on the left and $y$ to the $\act[b]$-labelled
event on the right).
Similarly, consider the formula $\psi = \en[a\,]{}{}{x}\en[b\,]{}{}{y}
\neg \en[c\,]{x,y}{}{z} \true$, saying that there are two events,
labelled by $\act$ and $\act[b]$, respectively, which are not common
causes for any $\act[c]$-labelled event. Also $\psi$ does not hold
neither in $\mathcal{E}_4$ nor in $\mathcal{E}_5$. Omitting the
restriction to legal pairs, $\psi$ would be true only in
$\mathcal{E}_4$ where $x$ and $y$ can be bound to conflicting events.
This means that the logic would allow one to distinguish the {\pes}s
corresponding to any process from that corresponding to the
non-deterministic choice between that process and itself, which
instead are equated by virtually any behavioural equivalence.

Instead of restricting the semantics of formulae to legal pairs, one
could envisage syntactic constraints which produce essentially the
same effect, thus limiting the observation power of the logic.  The
idea is quite simple: in any formula, whenever we bind an event to a
variable $z$, we require that the binder operator explicitly states the
consistency of $z$ with the free variables appearing in the remaining
part of the formula. Specifically, for any subformula of the kind
$\en{\vx}{\vy}{z}\, \psi$, we could require the free variables of
$\psi$ to be a subset of $\vx\cup\vy \cup \{z\}$. In this way we are
guaranteed that the event bound to $z$ is either causally dependent or
concurrent (hence consistent) with the events bound to the free
variables of the formula. This essentially gives the same effect as
restricting the semantics to legal pairs.
It can be seen that restricting to the fragment of $\lc$ consisting of
well-formed formulae does not alter the logical equivalence which
remains hhp-bisimilarity,  as for the full logic.
A more detailed account of this alternative approach is given in
Appendix~\ref{app:wf}.


\subsection{Dual operators}

Relying on negation we can define operators which are dual to those
primitive in the logic. As usual, disjunction $\varphi \lor \psi$ can
be defined by the formula $\neg (\neg \varphi \land \neg \psi)$. Its
semantics, according to Definition~\ref{de:semantics}, turns out to
be:
\begin{center}
  $\sem[\mathcal{E}]{\varphi \lor \psi} = (\sem[\mathcal{E}]{\varphi} \cup \sem[\mathcal{E}]{\psi}) \cap \ld{\varphi \lor \psi}$.
\end{center}
The formula $\false$ (false) is defined by $\neg \true$, with semantics:
\begin{center}
  $\sem[\mathcal{E}]{\false} = \emptyset$.
\end{center}
Moreover, we
write
 \begin{center}
 \begin{tabular}{rll}
   $\enBox{\vx}{\vy}{z}\, \varphi\ \ \ $ &
for the formula  & $\ \ \ \neg (\en{\vx}{\vy}{z}\, \neg \varphi)$.
\\[2mm]
 $\exBox{z}\, \varphi\ \ \ $ & for the formula & $\ \ \ \neg (\ex{z}\, \neg \varphi)$
\end{tabular}
 \end{center} 
The dual of the binder has a universal flavour. In fact its semantics, given explicitly below, involves a universal quantification:
 \begin{center}
   $
   \sem[\mathcal{E}]{\enBox{\vx}{\vy}{z}\, \varphi} \ = \
      \{ (C, \eta) \mid 
      \begin{array}[t]{ll}
           (C, \eta) \in \ld{\enBox{\vx}{\vy}{z}\, \varphi}\ \mathrm{and}\\
          \forall e \in \res{E}{C} \mbox{ such that }\ e \cons \eta(\fv{\varphi} \setminus \{ z \})\\
          \land\ 
           \lambda(e) = \act\ \land\ \eta(\vx) < e\ \land\ \eta(\vy) \conc e\\
           \mbox{it holds }  (C, \eta[z\mapsto e]) \in
           \sem[\mathcal{E}]{\varphi}  & \}\\[3mm]
      \end{array}
    $
\end{center}
i.e., $\mathcal{E}, C \models_\eta \enBox{\vx}{\vy}{z}\, \varphi$
when for all $\act$-labelled events $e$ in the future of $C$,
consistent with the events already bound to $\fv{\varphi}$, caused by
$\eta(\vx)$ and concurrent with $\eta(\vy)$, we have that binding $e$ to
$z$ the formula $\varphi$ holds.

The semantics of $\exBox{\cdot}$, instead, is:
\begin{center}
  $
  \sem[\mathcal{E}]{\exBox{z}\, \varphi} \ = \
  \{ (C, \eta) \mid 
  \begin{array}[t]{ll}
    (C, \eta) \in \ld{\exBox{z}\varphi}\ \mathrm{and}\\
    \mbox{if }C \trans{\eta(z)} C'\ 
    \mbox{ then }(C', \eta) \in
    \sem[\mathcal{E}]{\varphi} & \}
  \end{array}
  $
\end{center}
namely, $\mathcal{E}, C \models_\eta \exBox{z}\varphi$ if, either
$\eta(z)$ is not executable from $C$ or it is executable and in the
reached configuration $\varphi$ holds.

\medskip 

The logic $\lc$ could be alternatively defined in positive form by
including the dual operators and omitting negation. The
syntax of the resulting logic, denoted $\lc^+$, would be as follows:
\begin{center}
  $
  \varphi \ ::= \  
  \true \ \mid\ \false\ \mid\ 
  \varphi \land \varphi\ \mid\ \varphi\lor \varphi\ \mid\ 
  \en{\vx}{\vy}{z}\, \varphi\ \mid\ 
  \enBox{\vx}{\vy}{z}\,\varphi\ \mid\
  \ex{z}\varphi\ \mid\ \exBox{z}\,\varphi
  $
\end{center}

\noindent
Negation is then encodable in $\lc^+$ by duality.
%
%
%
Hereafter we will freely use the dual operators.

\subsection{Examples and notation}

In this subsection we provide some more examples illustrating the
expressiveness of the logic. We start by introducing some handy
notation, which will improve the readability of the formulae.

\paragraph{Immediate execution.}
We will write 
\begin{center}
  $\enx{\vx}{\vy}{z}\, \varphi\ \ \ $  for the formula   
  $\ \ \ \en{\vx}{\vy}{z} \ex{z} \varphi$
\end{center}
that states the existence of an event $e$ enabled by the current
configuration, and thus which can be immediately executed, such that
after executing $e$ the formula $\varphi$ holds (with $e$ bound to
variable $z$).
Dually we introduce the notation $\enxBox{\vx}{\vy}{z}\, \varphi$,
which stands for the formula $\enBox{\vx}{\vy}{z} \exBox{z} \varphi$.

\paragraph{Steps.}  
We introduce a notation also to predicate the existence, resp., the
immediate execution, of concurrent events, specifying also their
dependencies. We will write
\begin{center}
\begin{tabular}{rcl}
  $( \en{\vx}{\vy}{z} \sep \en[b]{\vx'}{\vy'}{z'} )\, \varphi\ \ $ &
  for the formula  &
$\ \ \en{\vx}{\vy}{z}\en[b]{\vx'}{\vy',z}{z'}\varphi $ 
\\[2mm]
  $ ( \enx{\vx}{\vy}{z} \sep \enx[b]{\vx'}{\vy'}{z'} ) \varphi
 \ \  $ & for the formula  &
$\ \ (\,
\en{\vx}{\vy}{z} \sep \en[b]{\vx'}{\vy'}{z'}\, )
\ex{z}\ex{z'}\varphi$
\end{tabular}
\end{center}
The first formula declares the existence of two concurrent events,
labelled by $\act$ and $\act[b]$, respectively, such that if we bind
such events to $z$ and $z'$, then $\varphi$ holds.
The second formula states the existence of two concurrently enabled
events, labelled by $\act$ and $\act[b]$, whose immediate execution leads to a
state where $\varphi$ holds.
In particular, the ability to perform a step
consisting of two concurrent events labelled by $\act$ and
$\act[b]$ is simply expressed by the formula
$(\enx{}{}{x}\sep\enx[b]{}{}{y})\true$.

Clearly, this notation can be generalised to the
quantification and the immediate execution of any number of concurrent
events.

An analogous notation will be used for the dual
operators:
\begin{center}
  $( \enBox{\vx}{\vy}{z} \sep \enBox[b]{\vx'}{\vy'}{z'} )\, \varphi$
   \qquad\mbox{ and }\qquad
  $ ( \enxBox{\vx}{\vy}{z} \sep \enxBox[b]{\vx'}{\vy'}{z'} ) \varphi
  $
\end{center}
The first formula asserts that considering any pair of concurrent
events, labelled $\act$ and $\act[b]$, respectively, which are bound
to $z$ and $z'$, the formula $\varphi$ holds. The second formula states that the after the execution of all pairs of concurrent events, labelled  
$\act$ and $\act[b]$, respectively, the formula $\varphi$ holds.

\begin{figure}[t]
\begin{center}
\begin{tabular}{cccc}
\mbox{
    \xymatrix@R=3mm@C=5mm{
      b  &  a    \\
      a\ar@{-}[u]\ar@{.}[r]  & b\ar@{-}[u]
    }}
&  \hspace{1cm}
\mbox{
    \xymatrix@R=3mm@C=4mm{
      \\
      a         &    b
    }}
&  \hspace{1cm}
 \mbox{
    \xymatrix@R=4mm@C=5mm{
      b  &  a    \\
      a\ar@{-}[u]\ar@{.}[ur] & b\ar@{-}[u]\ar@{.}[ul]   \\
    }}
&  \hspace{1cm}
\mbox{
    \xymatrix@R=3mm@C=4mm{
      b  &      \\
      a\ar@{-}[u]  & b\ar@{.}[ul] \\
    }}
\\ \\
$\mathcal{E}_6$
&  \hspace{1cm}
$\mathcal{E}_7$
&  \hspace{1cm}
$\mathcal{E}_8$
&  \hspace{1cm}
$\mathcal{E}_9$
\end{tabular}
\end{center}
  \caption{ }
  \label{fi:example2}
\end{figure}

\begin{example}[interleaving vs. true concurrency]
\label{ex:BisVSStep}
Consider the {\pes}s $\mathcal{E}_6$ and
$\mathcal{E}_7$ in Fig.~\ref{fi:example2}. They are equated by
interleaving equivalences and distinguished by any true concurrent
equivalence.
The formula $\varphi_1=\enx{}{}{x}\enx[b]{}{x}{y} \true = 
(\enx{}{}{x} \sep \enx[b]{}{}{y}) \true$ is true only on
$\mathcal{E}_7$, while 
$\varphi_2=\enx{}{}{x}\enx[b]{x}{}{y} \true$ is true only on
$\mathcal{E}_6$.
\end{example}

\paragraph{Wildcard operators.}
It is often useful to have a wildcard operator to refer to an event
with an arbitrary label. When the set of labels $\Lambda$ is finite,
we write
\begin{center}
  $\en[\_]{\vx}{\vy}{z}\varphi$
\end{center}
to denote the formula $\bigvee_{\act \in \Lambda}
\en[a]{\vx}{\vy}{z}\varphi$, and we use an analogous notation for the
induced operators.  For instance, the formula $(\enx[\_]{}{}{x_1} \sep
\enx[\_]{}{}{x_2})\true \,\wedge\, \neg ({\enx[\_]{}{}{y_1} \sep
  \enx[\_]{}{}{y_2}\sep\enx[\_]{}{}{y_3}) \true}$ states that in the
current state there is a step consisting of two concurrent events and
this is the maximal size for a step.
When the set of labels $\Lambda$ is infinite the same wildcard
operators are no longer expressible in the finitary logic $\lc$.  
However they
can be added to $\lc$ while retaining all the results in the paper.
More precisely, logical equivalence for $\lc$ would be still
hhp-bisimilarity. In fact, by adding the wildcard operators logical
equivalence becomes potentially finer and thus the fact that it
implies hhp-bisimilarity (Proposition~\ref{pr:LogicToHhp}) clearly
remains true. Conversely, finiteness of conjunctions plays no role in
the proof of Proposition~\ref{pr:HhpToLogic}, hence it can be easily
seen that hhp-bisimilarity implies logical equivalence even for an
infinitary version of the logic $\lc$ (explicitly introduced in
Section~\ref{ss:invariance} and denoted $\lc^\infty$) where wildcard
operators can be encoded.
The same applies to the various fragments of $\lc$ and to the logics
with recursion.
%
%


\begin{example}[causality and concurrency]
\label{ex:StepVSPomset}
Consider the {\pes}s $\mathcal{E}_6$ and $\mathcal{E}_8$ in
Fig.~\ref{fi:example2}. They are distinguished by all true concurrent
equivalences, but since they share the same causal structure, in order
to pinpoint how they differ, the logic must be able to express the
presence of two concurrent events. Logic $\lc$ can do this in a quite
direct way, e.g., $\mathcal{E}_8 \models (\enx{}{}{x} \sep
\enx[b]{}{}{y})\true$, while $\mathcal{E}_6 \not\models
(\enx{}{}{x} \sep \enx[b]{}{}{y})\true$.
On the other hand, {\pes}s $\mathcal{E}_7$ and $\mathcal{E}_9$,
roughly speaking, exhibit the same concurrency and indeed they are
equated by step bisimilarity. However they have a different causal
structure and thus they are distinguished by any equivalence which
observes causality, e.g., pomset bisimilarity. The logic can take them
apart by predicating directly about causality, e.g., $\mathcal{E}_9$
satisfies $\enx{}{}{x} \enx[b]{x}{}{y} \true$, while $\mathcal{E}_7$
does not.
\end{example}

\begin{example}[conflicting futures]
  \label{ex:hpVShhp}
  Consider the {\pes}s below which can be proved to be hp-bisimilar
  but not hhp-bisimilar (the example is taken from~\cite{JNW:BFOM}):
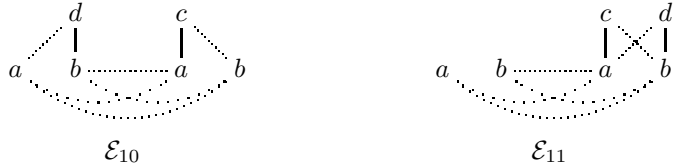
\begin{center}
\begin{tabular}{ccc}
\mbox{
\xymatrix@R=3mm@C=4mm{
              & d           &   &  c  & \\
a\ar@{.}[ur] \ar@/_1pc/@{.}[rrr]\ar@/_1.5pc/@{.}[rrrr] &
b\ar@{-}[u]\ar@{.}[rr]\ar@/_1pc/@{.}[rrr] &   & a \ar@{-}[u] & b\ar@{.}[ul] \\
}}
&\hspace*{2cm} &
\mbox{
\xymatrix@R=3mm@C=4mm{
       &     &    &  c  & d \\
    a \ar@/_1pc/@{.}[rrr]\ar@/_1.5pc/@{.}[rrrr] &  b \ar@{.}[rr]\ar@/_1pc/@{.}[rrr] &  &  a\ar@{-}[u]\ar@{.}[ur]  & b\ar@{-}[u]\ar@{.}[ul]
 }}\\[15mm]
$\mathcal{E}_{10}$ & & $\mathcal{E}_{11}$
\end{tabular}
\end{center}
Intuitively, they differ since the causes of the events labelled by $\act[c]$
and $\act[d]$, respectively, are in conflict in $\mathcal{E}_{10}$
and concurrent in $\mathcal{E}_{11}$. This difference can be captured by the formula
$\varphi=(\en{}{}{x}\sep
\en[b]{}{}{y})(\en[c]{x}{}{z_1}\true \wedge\en[d]{y}{}{z_2} \true)$,
which is satisfied only by $\mathcal{E}_{11}$.  Notice that the
formula $\varphi$ exploits the ability of the logic $\lc$ of
quantifying over events in conflict with previously bound events:
formula $\varphi$ is satisfied in $\mathcal{E}_{11}$ by binding $x$
and $y$ to the rightmost $\act$-labelled and $\act[b]$-labelled
events; then $z_1$ and $z_2$ are bound to events which are in
conflict with either $x$ or $y$. For this, the possibility of
``observing'' an event without executing it is essential: the
formula $\varphi'=(\enx{}{}{x} \sep \enx[b]{}{}{y})
(\en[c]{x}{}{z_1}\true \wedge \en[d]{y}{}{z_2} \true)$ would
be false for both {\pes}s since the execution of the first two events
leads to a configuration that is no further extensible.

As a last example, consider the CCS processes $P=a|(b\!+\!c) + a|b
+ b|(a\!+\!c)$ and $Q=a|(b\!+\!c) + b|(a\!+\!c)$, equated by the absorption law (see, e.g.,~\cite{vGlab01}).  They contain no
causal dependencies, but they exhibit a different interplay between
concurrency and branching. Accordingly, the corresponding {\pes}s can
be proved to be hp-bisimilar but not hhp-bisimilar.  Intuitively, this
difference arises from the fact that only the process $P$ includes two
concurrent events $\act$ and $\act[b]$ such that, once their execution has started, by firing one of them, no $\act[c]$-labelled event will ever be
enabled. Such a difference can be expressed in $\lc$ by the formula
$(\en{}{}{x} \sep \en[b]{}{}{y}) (\neg {\en[c]{}{x}{z} \true} \wedge
\neg {\en[c]{}{y}{z'}\true})$, which says that there are two
concurrent events labelled $\act[a]$ and $\act[b]$, respectively, such
that none of them is concurrent with a $\act[c]$-labelled event. This
is clearly satisfied only by the {\pes} corresponding to $P$.
\end{example}

\section{A logical characterisation of hhp-bisimilarity}
\label{se:hhp-characterisation}

We next study the logical equivalence induced by $\lc$. We have
already argued that no formula in $\lc$ distinguishes the {\pes}s $a$
and $a \# a$, hence the logical equivalence induced by $\lc$ is surely
coarser than isomorphism. In this section we will show that it
coincides with hhp-bisimilarity.

Since later we will also identify suitable fragments of $\lc$
corresponding to coarser equivalences, we define logical equivalence
for a generic fragment of $\lc$.

\begin{definition}[logical equivalence] 
  Let $\mathcal{L}'$ be a fragment of $\lc$. We say that two {\pes}
  $\mathcal{E}_1,\mathcal{E}_2$ are \emph{logically equivalent} in
  $\mathcal{L}'$, written
  $\mathcal{E}_1\equiv_{\mathcal{L}'}\mathcal{E}_2$ when they satisfy
  the same closed formulae of $\mathcal{L}'$.
\end{definition}

We first prove that two {\pes}'s satisfying the same formulae in $\lc$
are hhp-bisimilar.

\begin{proposition}
  \label{pr:LogicToHhp}
  Let $\mathcal{E}_1$ and $\mathcal{E}_2$ be {\pes}s such
  that $\mathcal{E}_1 \equiv_{\lc} \mathcal{E}_2$, then $\mathcal{E}_1
  \hhpb \mathcal{E}_2$.
\end{proposition}

\begin{proof}
  Let us start by introducing some notation. We fix a surjective
  environment $\eta_1 : \Var \to E_1$. Then given an event $e \in
  E_1$, we write $x_{e}$ to denote a fixed distinguished variable
  such that $\eta_1(x_{e}) = e$. Similarly, for a configuration
  $C_1 = \{e_1,\ldots,e_n\}$ we denote by $X_{C_1}$ the set of
  variables $\{ x_{e_1}, \ldots, x_{e_n}\}$. Observe that
  $(\emptyset,\eta_1)$ is a legal pair for any formula $\varphi\in\lc$
  such that $\fv{\varphi}\subseteq X_{C_1}$, since $\emptyset \cup
  \eta(\fv{\varphi}) \subseteq C_1$, which is consistent.
  
  \noindent
  Consider the posetal relation $R \subseteq
  \conf{\mathcal{E}_1} \bar{\times} \conf{\mathcal{E}_2}$ defined by:
  \begin{equation}
    \label{eq:relation}
    R \ =\ \{\, (C_1,f,C_2) \mid \forall \psi \in \lc.\ 
    \fv{\psi}\subseteq X_{C_1}\ 
    \ \ (\mathcal{E}_1, \emptyset  \models_{\eta_1}\psi \
    \mathit{iff}\ \mathcal{E}_2, \emptyset \models_{f \circ \eta_1}\psi)
    \, \}
  \end{equation}
  where, for an isomorphism of pomsets $f : C_1 \to C_2$, we denote by
  $f \circ \eta_1$ an environment such that $f \circ \eta_1(x) =
  f(\eta_1(x))$ for $x \in X_{C_1}$ and $f \circ \eta_1(x)$ has any
  value, otherwise. Note that this does not introduce ambiguities,
  since, by Lemma~\ref{le:env-fv}, the semantics of $\psi$ only
  depends on the value of the environment on $\fv{\psi}$ and
  $\fv{\psi} \subseteq X_{C_1}$ by construction.
  
  \smallskip

  Observe that, since by hypothesis $\mathcal{E}_1 \equiv_{\lc}
  \mathcal{E}_2$, we have that $(\emptyset, \emptyset, \emptyset) \in
  R$. Hence in order to conclude it is sufficient to show that $R$ is
  a hhp-bisimulation.
  \begin{itemize}
  \item ${R}$ is downward closed\\
    Take $(C_1,f,C_2)\in R$ and consider
    $(C'_1,f',C'_2)\subseteq (C_1,f,C_2)$ pointwise. We have to show
    that $(C'_1,f',C'_2)\in R$.

    Let $\psi$ be any formula such that $\fv{\psi}\subseteq X_{C_1'}$.
    Since $C_1' \subseteq C_1$, clearly $\fv{\psi} \subseteq X_{C_1}$
    and thus, since $(C_1,f,C_2)\in R$, by definition of $R$
    (\ref{eq:relation}), we have that
    \begin{center}
      $\mathcal{E}_1, \emptyset \models_{\eta_1}\psi$
      \quad iff \quad 
      $\mathcal{E}_2, \emptyset \models_{f \circ \eta_1}\psi$, 
    \end{center}
    Moreover, since $\fv{\psi} \subseteq X_{C_1'}$, $\eta_1(X_{C_1'})
    = C_1'$ and $f' = f_{|C_1'}$, we have that $(f \circ
    \eta_1)_{|\fv{\psi}} = (f' \circ \eta_1)_{|\fv{\psi}}$ and thus by
    Lemma~\ref{le:env-fv},
    \begin{center}
      $\mathcal{E}_2, \emptyset \models_{f \circ \eta_1}\psi$ 
      \qquad iff \qquad
      $\mathcal{E}_2, \emptyset \models_{f' \circ \eta_1}\psi$
    \end{center}
    Summing up, for any $\psi$ such that $\fv{\psi} \subseteq X_{C_1'}$, it
    holds that $\mathcal{E}_1, \emptyset \models_{\eta_1}\psi$ iff
    $\mathcal{E}_2, \emptyset \models_{f' \circ
      \eta_1}\psi$. Therefore $(C_1', f', C_2') \in R$, as desired.
    
     \medskip
    
  \item $R$ is a hp-bisimulation\\ 
    We have to show that given $(C_1,f,C_2)\in R$, if $C_1\trans{e}
    C_1'$ then there exists a transition $C_2\trans{g} C_2'$ such that
    $f'=f[e\mapsto g] : C_1' \to C_2'$ is an isomorphism of pomsets
    (hence in particular $\lambda_1(e)=\lambda_2(g)$) and
    $(C_1',f',C_2')\in R$.

    We proceed by contradiction. Since all {\pes}s are assumed to be
    image finite, there are finitely many transitions $C_2 \trans{g^i}
    C_2^i$, with $i \in \{1, \ldots, n\}$, such that $C_1' \sim C_2^i$
    (as pomsets). By contradiction assume that, for any $i \in \{ 1,
    \ldots, n\}$, it holds $(C_1',f^i,C_2^i) \not\in R$. Hence, by
    definition of $R$ (\ref{eq:relation}), there exists a formula
    $\psi^i$ such that
    \begin{center}
      $\mathcal{E}_1, \emptyset  \models_{\eta_1} \psi^i$ 
      \quad and \quad
      $\mathcal{E}_2, \emptyset \not\models_{f^i\circ\eta_1}\psi^i$
    \end{center}
    where $\fv{\psi^i}\subseteq X_{C_1'}=X_{C_1}\cup\{x_e\}$
    and $f^i=f[e\mapsto g^i]$. 
    Observe that it could either be that $\mathcal{E}_1, \emptyset
    \not\models_{\eta_1} \psi^i$ and $\mathcal{E}_2, \emptyset
    \models_{f^i\circ\eta_1}\psi^i$, but we can reduce to the case
    above by taking the negation of $\psi^i$. In fact, since $\fv{\psi^i}
    \subseteq X_{C_1'}$, we have that $(\emptyset, \eta_1) \in
    \ld[\mathcal{E}_1]{\psi^i}$, and thus from $\mathcal{E}_1, \emptyset
    \not\models_{\eta_1} \psi^i$ we deduce $\mathcal{E}_1, \emptyset
    \models_{\eta_1} \neg\psi^i$. Moreover, since $\mathcal{E}_2,
    \emptyset \models_{f^i\circ\eta_1}\psi^i$ we have $\mathcal{E}_2,
    \emptyset \not\models_{f^i\circ\eta_1} \neg\psi^i$.

    \smallskip

    Consider the formula
    \begin{center}
      $\varphi = \en{\vx}{\vy}{x_e}(\ex{X_{C_1}} \ex{x_e} \true
     \land \psi^1\land\ldots\land\psi^n)$
    \end{center}
    where $\act = \lambda_1(e)$ and the $\vx, \vy \subseteq
    X_{C_1}$ are such that $\eta_1(\vx)$ is the set of causes of $e$
    in $C_1$ and $\eta_1(\vy)$ is the set of events in $C_1$ which are
    concurrent with $e$.  Note that 
    \begin{center}
      $\fv{\varphi} = \vx \cup \vy \cup ((X_{C_1} \cup \{x_e\}  \cup
    \bigcup_{i=1}^n \fv{\psi_i}) \setminus \{ x_e \}) =
    X_{C_1}$
    \end{center}
    In fact, by construction, $\vx \cup \vy = X_{C_1}$  and 
    $\fv{\psi^i}\subseteq X_{C_1'}=X_{C_1}\cup\{x_e\}$.

    Now, it is easy to see that $\mathcal{E}_1, \emptyset
    \models_{\eta_1} \varphi$. Moreover $\mathcal{E}_2,\emptyset
    \not\models_{f \circ \eta_1}\varphi$. In fact, an event $g \in
    E_2$ such that $f \circ \eta_1(\vx) < g$, $f \circ \eta_1(\vy)
    \conc g$ and $\mathcal{E}_2,\emptyset \models_{f \circ \eta_1}
    \ex{X_{C_1}} \ex{x_e}$ is necessarily in the set $\{ g^1, \ldots,
    g^n\}$, and thus, by construction, $\mathcal{E}_2,\emptyset
    \not\models_{f \circ \eta_1[x_e \mapsto g]} \psi^i$ for some $i
    \in \{1, \ldots, n\}$. 

    The existence of a formula $\varphi$ which distinguishes $C_1$ and
    $C_2$ contradicts the hypothesis $(C_1, f, C_2) \in R$, as desired.
    
    \smallskip

    The fact that also the converse holds, i.e., if $C_2\trans{g}
    C_2'$ then there exists a transition $C_1\trans{e} C_1'$ such that
    $f'=f[e\mapsto g] : C_1' \to C_2'$ is an isomorphism of pomsets
    and $(C_1',f',C_2')\in R$, can be proved analogously.  \qed
  \end{itemize}
\end{proof}

In order to prove that, conversely, hhp-bisimilar {\pes}s satisfy the
same $\lc$ formulae, we first recall a lemma
from~\cite{Ber:HHP,vGlab01} which will be useful in the sequel.

\begin{lemma}[hhp-bisimilarity as a {\pes}]
  \label{le:span}
  Let $\mathcal{E}_1$, $\mathcal{E}_2$ be {\pes}s such that
  $\mathcal{E}_1 \hhpb \mathcal{E}_2$ and let $R$ be a
  hhp-bisimulation. Then there exists a {\pes} $\mathcal{E}_{R} =
  \langle E_{R}, \leq_{R}, \#_{R}, \lambda_{R} \rangle$ such that for
  $i \in \{ 1,2\}$
  \begin{itemize}

  \item $\mathcal{E}_i \hhpb \mathcal{E}_{R}$

  \item there are surjective maps $f^i_{R}:E_{R}\to
    E_i$ such that 
    $\{\,(C,{f^i_R}_{|C},f^i_{R}(C)) \mid C \in \conf{\mathcal{E}_R} \}$ is
    a hhp-bisimulation.
  \end{itemize}
  Additionally, each $f^i_{R}$ preserves labels, $\leq$ and $\conc$, maps
  configurations to configurations and it is injective on
   consistent sets of events.
\end{lemma}

\begin{proof}[Sketch, from \cite{Ber:HHP,vGlab01}]
  We just recall the definition of $\mathcal{E}_{R} = \langle E_{R},
  \leq_{R}, \#_{R}, \lambda_{R} \rangle$:
 \begin{itemize}
  \item $E_{R}=\{(e_1,f,e_2) \mid (\cause{e_1}, f, \cause{e_2}) \in R\}$,
  \item $(e_1, f, e_2) \leq_{R} (e_1', f', e_2')$ if
    $f \subseteq f'$,
  \item $(e_1, f, e_2) \#_{R}  (e_1', f', e_2')$ if there exists no $(C,g,D)\in{R}$ such that
    $(\cause{e_1}, f, \cause{e_2}), (\cause{e_1'},f', \cause{e_2'})
    \subseteq (C,g,D)$ pointwise,
  \item $\lambda_{R}(e_1,f,e_2)= \lambda_1(e_1)$.
\end{itemize}
The maps $f^1_{R}:E_{R}\to E_1$ and $f^2_{R}:E_{R}\to E_2$ are
just the projections on the first and third components, respectively.
\qed
\end{proof}

\begin{proposition}
  \label{pr:HhpToLogic}
  Let $\mathcal{E}_1$ and $\mathcal{E}_2$ be {\pes}s such that
  $\mathcal{E}_1 \hhpb \mathcal{E}_2$. Then $\mathcal{E}_1
  \equiv_{\lc} \mathcal{E}_2$.
\end{proposition}

\begin{proof}
  Let $R$ be a hhp-bisimulation relating $\mathcal{E}_1$ and
  $\mathcal{E}_2$. By Lemma~\ref{le:span}, it is not restrictive to
  assume that $R = \{\, (C_1,f_{|_{C_1}},f(C_1))\, \}$, where
  $f:E_1\to E_2$ is a surjective map satisfying the conditions in the
  statement of the lemma.
  Then it is sufficient to prove that 
  for any formula $\varphi \in \lc$, for
  any $(C_1,\eta_1) \in \ld[\mathcal{E}_1]{\varphi}$
  \begin{equation}
    \label{eq:hhp-log}
    \mathcal{E}_1, C_1 \models_{\eta_1} \varphi 
    \quad \mathrm{iff} \quad
    \mathcal{E}_2,f(C_1) \models_{f \circ \eta_1} \varphi
  \end{equation}
  This implies, in particular, that $\mathcal{E}_1$ and
  $\mathcal{E}_2$ satisfy the same closed formulae, i.e.,
  $\mathcal{E}_1 \equiv_{\lc} \mathcal{E}_2$ as desired. In fact,
  given any closed formula $\varphi$, note that $(\emptyset, \eta_1) \in
  \ld[\mathcal{E}_1]{\varphi}$ for all environments $\eta_1$. Therefore
  if $\mathcal{E}_1 \models \varphi$, which means $\mathcal{E}_1,
  \emptyset \models_{\eta_1} \varphi$ for some $\eta_1$, we have
  $\mathcal{E}_2, \emptyset \models_{f \circ \eta_1} \varphi$, i.e.,
  $\mathcal{E}_2 \models \varphi$. Vice versa, if $\mathcal{E}_2
  \models \varphi$ then $\mathcal{E}_2, \emptyset \models_{\eta_2}
  \varphi$ for some $\eta_2 \in \Env_{\mathcal{E}_2}$. Since $\varphi$
  is closed, by Lemma~\ref{le:env-fv} the environment is irrelevant
  and thus, if we take any $\eta_1 \in \Env_{\mathcal{E}_1}$, it holds
  $\mathcal{E}_2, \emptyset \models_{f \circ \eta_1} \varphi$. By this
  we get $\mathcal{E}_1, \emptyset \models_{\eta_1} \varphi$, which means
  $\mathcal{E}_1 \models \varphi$.

  Now, in order to prove (\ref{eq:hhp-log}), first of all note that
  $f$ preserves legal pairs, i.e., if $(C_1, \eta_1) \in
  \ld[\mathcal{E}_1]{\varphi}$ then $(f(C_1), f \circ \eta_1) \in
  \ld[\mathcal{E}_2]{\varphi}$ since $f$ preserves consistency (as it
  preserves causality and concurrency).

  The proof proceeds by induction on the formula $\varphi$:

   \begin{itemize}

   \item $\varphi=\true$\\
     Immediate.

     \medskip

   \item $\varphi = \varphi_1\land\varphi_2$\\
     Let $(C_1, \eta_1) \in
     \ld[\mathcal{E}_1]{\varphi}$, hence $(C_1, \eta_1) \in
     \ld[\mathcal{E}_1]{\varphi_i}$ for $i \in \{1,2\}$.
     If $\mathcal{E}_1, C_1 \models_{\eta_1} \varphi$, then, by
     definition of the semantics, we have $\mathcal{E}_1, C_1
     \models_{\eta_1} \varphi_i$, for $i \in \{ 1,2\}$.  Thus we can
     use the inductive hypothesis to get that $\mathcal{E}_2, f(C_1)
     \models_ {f \circ \eta_1} \varphi_i$, for $i \in \{
     1,2\}$. Moreover, since $f$ preserves legal pairs, we know that
     $(f(C_1), f \circ \eta_1) \in
     \ld[\mathcal{E}_2]{\varphi}$. Therefore $\mathcal{E}_2, f(C_1)
     \models_{f \circ \eta_1} \varphi$. The converse implication can
     be proved by just reverting all deductions.

     \medskip

   \item 
     $\varphi = \neg\varphi_1$\\
     Analogous to the previous case.

     \medskip
     
   \item 
     $\varphi=\en{\vx}{\vy}{z}\psi$\\
     Assume that $\mathcal{E}_1, C_1 \models_{\eta_1} \varphi$, with
     $(C_1, \eta_1) \in \ld[\mathcal{E}_1]{\varphi}$.  Hence, by
     definition of the semantics, there exists an event $e \in
     \res{E_1}{C_1}$, such that $e \cons \eta_1(\fv{\psi} \setminus \{ z
     \})$, $\lambda_1(e) = \act$, $\eta_1(\vx) \leq e$, $\eta_1(\vy)\conc
     e$ and
     \begin{equation}
       \label{eq:hhp1}
       \mathcal{E}_1, C_1 \models_{\eta_1'} \psi
     \end{equation}
     where $\eta_1' = \eta_1[z \mapsto e]$.
     
     By (\ref{eq:hhp1}) and Lemma~\ref{le:legal}, $(C_1,\eta_1') \in
     \ld[\mathcal{E}_1]{\psi}$. Hence by inductive hypothesis
     $\mathcal{E}_2, f(C_1) \models_{f \circ \eta_1'} \psi$, with $f
     \circ \eta_1' = (f \circ \eta_1)[z\mapsto f(e)]$.
     
     Since, by Lemma~\ref{le:span}, $f$ preserves consistency and it
     is injective on consistent sets of events, $f(e) \in
     \res{E_2}{f(C_1)}$.
     Additionally, again by Lemma~\ref{le:span}, since $f$ preserves
     labels, $\leq$ and $\conc$ (and hence $\cons$) we have that
     $f(e) \cons f \circ \eta_1(\fv{\psi} \setminus \{ z \})$, 
     $\lambda_2(f(e)) = \lambda_1(e) = \act$ and $f(\eta_1(\vx)) \leq
     f(e)$, $f(\eta_1(\vy)) \conc f(e)$.
     Therefore we conclude that, as desired
     \begin{center}
       $\mathcal{E}_2, f(C_1) \models_{f \circ \eta_1} \varphi$.
     \end{center}

     \bigskip
     
     Conversely, let $\mathcal{E}_2, f(C_1) \models_{f \circ \eta_1}
     \varphi$, where $(C_1, \eta_1) \in
     \ld[\mathcal{E}_1]{\varphi}$. Therefore there exists an event $g
     \in \res{E_2}{f(C_1)}$, such that $g \cons f \circ \eta_1(\fv{\psi}
     \setminus \{ z \})$,
     $\lambda_2(g) = \act$,
     $f(\eta_1(\vx))\leq g$ and $f(\eta_1(\vy)) \conc g$ and
     $\mathcal{E}_2, f(C_1) \models_{\eta_2'} \psi$, where $\eta_2' = (f
     \circ \eta_1)[z \mapsto g]$.

     From the fact that $\mathcal{E}_2, f(C_1) \models_{f \circ \eta_1}
     \varphi$, by Lemma~\ref{le:legal}, we have that $(f(C_1), f \circ
     \eta_1) \in \ld[\mathcal{E}_2]{\varphi}$.  This means that $f(C_1)
     \cup f \circ \eta_1(\fv{\varphi})$ is consistent and thus $D_2 =
     f(C_1) \cup \cause{f \circ \eta_1(\fv{\varphi})}$ is a
     configuration. Since $\fv{\varphi} = \vx \cup \vy \cup (\fv{\psi}
     \setminus \{ z \})$, the arguments above show that
     \begin{equation}
       \label{eq:g-cons}
       D_2 \cons g.
     \end{equation}
     
     Now, since by hypothesis $(C_1,\eta_1) \in
     \ld[\mathcal{E}_1]{\varphi}$, we know that $C_1 \cup
     \eta_1(\fv{\varphi})$ is consistent. It follows that $D_1 = C_1
     \cup \cause{\eta_1(\fv{\varphi})}$
     is a configuration. Since, by Lemma~\ref{le:span}, $f$ is
     injective on consistent sets and preserves causality,
     \begin{quote}
       \begin{tabular}{lll}
         $D_2$ & $=$ & $f(C_1) \cup \cause{f \circ \eta_1(\fv{\varphi})}$\\
               & $=$ & $f(C_1) \cup f(\cause{\eta_1(\fv{\varphi})}$\\
               & $=$ & $f(C_1 \cup \cause{\eta_1(\fv{\varphi})})$\\
               & $=$ & $f(D_1)$
       \end{tabular}
     \end{quote}
     which means that $(D_1, f_{|D_1}, D_2) \in R$.

     \smallskip

     We distinguish two cases. If $g \in D_2$, since $f_{|D_1}$
     is an isomorphism of pomsets between $D_1$ and $D_2$, we can take
     the (unique) $e \in D_1$ such that $f(e) = g$. By using the
     isomorphism property, we have immediately that $e \in
     \res{E_1}{C_1}$, $\eta_1(\fv{\psi} \setminus \{ z \}) \cons e$,
     $\lambda_1(e) = \lambda_2(g)=\act$, $\eta_1(\vx)\leq e$ and
     $\eta_1(\vy) \conc e$.
     Define the environment $\eta_1' = \eta_1[z \mapsto e]$. Note that
     $(C_1, \eta_1')  \in \ld[\mathcal{E}_1]{\psi}$ since $C_1 \cup
     \eta_1'(\fv{\psi}) \subseteq C_1 \cup \eta_1'(\fv{\varphi} \cup \{ z
     \}) \subseteq D_1$. Therefore, since $\mathcal{E}_2, f(C_1)
     \models_{\eta_2'} \psi$, noticing that $f \circ \eta_1' = \eta_2'$,
     by inductive hypothesis we conclude $\mathcal{E}_1, C_1
     \models_{\eta_1'} \psi$. Hence
     \begin{center}
       $\mathcal{E}_1, C_1 \models_{\eta_1} \varphi$
     \end{center}

     \smallskip

     Otherwise, if $g \not\in D_2$, recalling (\ref{eq:g-cons}), if we
     let $X_2 = \cause{g} \setminus D_2$ we have a pomset transition in $\mathcal{E}_2$:
     \begin{equation}
       \label{eq:trans2}
       D_2 \trans{X_2} D_2'
     \end{equation}
         
     Therefore, since $R$ is a hhp-bisimulation, there is a pomset
     transition in $\mathcal{E}_1$ simulating (\ref{eq:trans2}):

     \begin{equation}
       \label{eq:trans1}
       D_1 \trans{X_1} D_1'
     \end{equation}
     such that $(D_1', f_{|D_1'}, D_2') \in R$.
     Now, $g \in D_2'$ and thus we can replicate the argument above.
     
     \medskip

   \item $\varphi=\ex{x}\psi$\\
     Assume that $\mathcal{E}_1, C_1 \models_{\eta_1} \varphi$, where
     $(C_1,\eta_1) \in \ld[\mathcal{E}_1]{\varphi}$.
     By definition of the semantics this means that
     \begin{center}
       $C_1 \trans{\eta_1(x)} C_1'$
     \end{center}
     and $\mathcal{E}_1, C_1' \models_{\eta_1} \psi$.
     
     Since $R$ is a hhp-bisimulation, we have that
     \begin{center}
       $f(C_1) \trans{f(\eta_1(x))} f(C_1')$.
     \end{center}
     Now, since $C_1' = C_1 \cup \{\eta_1(x)\}$ and $\fv{\psi} \subseteq
     \fv{\varphi}$, we have that
     \begin{center}
       $C_1' \cup \eta_1(\fv{\psi}) \subseteq C_1 \cup \{\eta_1(x)\} \cup
       \eta_1(\fv{\varphi}) = C_1 \cup \eta_1(\fv{\varphi})$.
     \end{center}
     Since $(C_1, \eta_1) \in \ld[\mathcal{E}_1]{\psi}$ the set above is
     consistent and thus $(C_1', \eta_1) \in
     \ld[\mathcal{E}_1]{\psi}$. Therefore we can use the inductive
     hypothesis to deduce $\mathcal{E}_2, f(C_1') \models_{f \circ
       \eta_1} \psi$ and thus, as desired,
     \begin{center}
       $\mathcal{E}_2, f(C_1) \models_{f \circ \eta_1} \varphi$.
     \end{center}
     
     \bigskip

     Conversely, let $\mathcal{E}_2, f(C_1) \models_{f \circ \eta_1}
     \varphi$, where $(C_1,\eta_1) \in \ld[\mathcal{E}_1]{\varphi}$.
     By definition of the semantics this means that
     \begin{center}
       $f(C_1) \trans{f(\eta_1(x))} C_2'$
     \end{center}
     and $\mathcal{E}_2, C_2' \models_{f \circ \eta_1} \psi$.
     
     Since $(C_1,\eta_1) \in \ld[\mathcal{E}_1]{\psi}$, we know that
     $\eta_1(x)$ is consistent with $C_1$.
     Moreover, $C_1 \cup \{ \eta_1(x) \}$ is causally closed,
     otherwise, since $f$ preserves causality and it is injective on
      consistent sets, also $f(C_1 \cup \eta_1(x)) = C_2 \cup
     f(\eta_1(x)) = C_2'$ would not be causally closed.
     
     Hence $C_1' = C_1 \cup \{ \eta_1(x) \}$ is a configuration and thus
     \begin{center}
       $C_1 \trans{\eta_1(x)} C_1'$
     \end{center}
     and clearly $f(C_1') = C_2'$.
     As above we can show that $(C_1', \eta_1) \in
     \ld[\mathcal{E}_1]{\psi}$ and thus, by inductive hypothesis,
     $\mathcal{E}_1, C_1' \models_{\eta_1} \psi$. Hence, as desired
     \begin{center}
       $\mathcal{E}_1, C_1 \models_{\eta_1} \varphi$.
     \end{center}
   \end{itemize}
   \qed
\end{proof}

Propositions~\ref{pr:HhpToLogic} and~\ref{pr:LogicToHhp} together say
that hhp-bisimilarity is the logical equivalence of $\lc$.

\begin{theorem}[hhp-bisimilarity]
  \label{th:hhp-bisimilarity}
  Let $\mathcal{E}_1$ and $\mathcal{E}_2$ be {\pes}s. Then
  $\mathcal{E}_1 \hhpb \mathcal{E}_2$ iff $\mathcal{E}_1 \equiv_{\lc}
  \mathcal{E}_2$.
\end{theorem}

\section{From Hennessy-Milner logic to HP-logic}
\label{se:spectrum}

Hhp-bisimilarity is the finest equivalence in the spectrum of true
concurrent equivalences proposed in~\cite{vGlab01}. Interestingly
enough, coarser equivalences such as step, pomset and hp-bisimilarity,
can be captured by suitable fragments of $\lc$ summarised in
Fig.~\ref{fig:Fragments}, which can be viewed as the logical
counterpart of the true concurrent spectrum.

Note that in each of these fragments after predicating the existence
of an event we must execute it. As a consequence, differently from
what happens in the full logic, in the fragments it is impossible to
refer to events in conflict with already observed events. Intuitively,
this says that behavioural equivalences up to hp-bisimilarity can observe
events only by executing them. Hence they cannot fully capture the
interplay between concurrency and branching, which is indeed
distinctive of hhp-bisimilarity.

\begin{figure}[ht]
\begin{center}
\begin{tabular}{c@{\hspace{3mm}}l@{\hspace{3mm}}l}
HM Logic &  $\lchm$ &
  $\varphi\ ::=\ \enx{}{}{x}\, \varphi\ \mid\ \varphi \land
  \varphi\ \mid\ \neg \varphi\ \mid\ \true$
\\[4mm]
Step Logic & $\lcs$ &
  $\varphi\ ::=\ (\enx[a_1]{}{}{x_1}\sep \cdots \sep\enx[a_n]{}{}{x_n})\ \varphi\ \mid\ \varphi
    \land \varphi\ \mid\ \neg  \varphi\ \mid\ \true$
\\[4mm]
Pomset Logic & $\lcp$ &
  $\varphi\ ::=\ \enx{\vx}{\vy}{z}\,  \varphi\ \mid\
    \neg \varphi\ \mid\ \varphi \land \varphi\ \mid\ \true$\\
    & & where $\neg$, $\land$ are used only on closed formulae.
\\[4mm]
HP Logic & $\lchp$ &
  $\varphi\ ::=\ \enx{\vx}{\vy}{z}\,  \varphi\ \mid\
    \neg \varphi\ \mid\ \varphi \land \varphi\ \mid\ \true$
\end{tabular}
\end{center}
\caption{Fragments of $\lc$ corresponding to various behavioural equivalences}
\label{fig:Fragments}
\end{figure}

\subsection{Hennessy-Milner logic}

A first simple observation is that standard Hennessy-Milner logic can be
recovered as the fragment of $\lc$ where only the derived
modality $\enx{}{}{x}\varphi$ (with no references to causally
dependent/concurrent events) is allowed.
In words, whenever we state the existence of an event we are forced to
execute it. Note that, since no dependencies can be expressed, the
bound variable $x$ is irrelevant.  The induced logical equivalence is
thus (interleaving) bisimilarity~\cite{HM:ALCND} (recall that we
consider only image finite {\pes}'s).

\subsection{Step logic}

A fragment $\lcs$ corresponding to step bisimilarity naturally arises
as a generalisation of HM logic where we can refer to sets of
concurrently enabled events. More precisely, as shown in
Fig.~\ref{fig:Fragments},
%
%
$\lcs$ is the fragment of $\lc$ where only the derived modality
$\enx[a_1]{}{}{x_1}\sep \cdots \sep\enx[a_n]{}{}{x_n}$ is used,
allowing to predicate on the possibility of performing a parallel
step, but without any reference to causal dependencies. Note that all
formulae in $\lcs$ are closed, and thus environments (as well as
variables) are irrelevant in their semantics.

As an example, consider the two {\pes}s $\mathcal{E}_6$ and
$\mathcal{E}_7$ in Fig.~\ref{fi:example2}. 
They are bisimilar but not step bisimilar since only $\mathcal{E}_7$  can
execute the step consisting of $\act$ and $\act[b]$ in parallel. Accordingly, they are taken apart by the
formula $(\enx[a]{}{}{}\sep\enx[b]{}{}{}) \true$ in $\lcs$, which is true only
on $\mathcal{E}_7$.

\begin{lemma}
  \label{le:step-logic}
  Let $\mathcal{E}_1$ and $\mathcal{E}_2$ be {\pes}s and let $C_i \in
  \conf{\mathcal{E}_i}$, for $i \in \{ 1,2 \}$, be
  configurations. There exists a step bisimulation $R$ such that
  $(C_1, C_2) \in R$ iff for any $\varphi \in \lcs$, $\mathcal{E}_1,
  C_1 \models \varphi \Leftrightarrow \mathcal{E}_2, C_2 \models
  \varphi$.
\end{lemma}

\begin{proof}
  ($\Rightarrow$) Assume that $(C_1,C_2) \in R$ for some step
  bisimulation $R$.
  The proof that for all $\varphi\in\lcs$, $\mathcal{E}_1,C_1
  \models \varphi$ iff $\mathcal{E}_2,C_2\models \varphi$ can be carried out
  by induction on the structure of $\varphi$. 

  We only discuss the non-trivial case where
  $\varphi=(\enx[a_1]{}{}{x_1}\sep\cdots
  \sep\enx[a_n]{}{}{x_n})\ \psi$. Assume that $\mathcal{E}_1,C_1
  \models \varphi$. Hence there is a step $C_1 \trans{\{e_1, \ldots,
    e_n\}} C_1'$ where $\lambda_1(e_i)=\act_i$ for $i \in \{1,
  \ldots,n\}$ and
  \begin{equation}
    \label{eq:step1}
    \mathcal{E}_1,C_1' \models \psi.
  \end{equation}
  Since $(C_1, C_2) \in R$, also $C_2$ can perform an
  analogous step
  \begin{center}
    $C_2 \trans{\{g_1, \ldots, g_n\}} C_2'$
  \end{center}
  with $\lambda_2(g_i)=\act_i$ for $i \in \{1,\ldots,n\}$ and
  $(C_1', C_2') \in R$. Additionally, by (\ref{eq:step1}) and the
  induction hypothesis, we have that $\mathcal{E}_2,C'_2 \models
  \psi$. Therefore we conclude $\mathcal{E}_2,C_2 \models \varphi$.

  \bigskip

  ($\Leftarrow$) We prove that the relation 
  \begin{center}
    $R=\{(C_1,C_2) \mid \forall \varphi\in\lcs\ \ (\mathcal{E}_1,C_1
        \models \varphi$ iff $\mathcal{E}_2,C_2\models \varphi) \}$
  \end{center}
  is a step bisimulation.

  We proceed by contradiction. Let $(C_1,C_2) \in R$, let $C_1
  \trans{X} C_1'$ be a step in $\mathcal{E}_1$ and assume that for all $Y$
  such that $C_2 \trans{Y} C_2'$ and $X \sim Y$ as pomsets it does not
  hold that $(C_1', C_2') \in R$. Hence there exists a formula $\psi \in \lcs$
  such that $\mathcal{E}_1, C_1' \models \psi$ and $\mathcal{E}_2,C_2'
  \not\models \psi$.
  
  Since our {\pes}s are assumed to be image finite, the
  number of possible steps $C_2 \trans{Y} C_2'$, with $X \sim Y$ is
  finite. Let $C_2\trans{Y^i}C_2^i$, for $i \in \{1, \ldots, k\}$, be such
  steps and let $\psi^i$ be the formulae such that $\mathcal{E}_1,C'_1
  \models \psi^i$ and $\mathcal{E}_2,C_2^i \not\models \psi^i$.  If we define
  \begin{center}
    $\psi = (\enx[a_1]{}{}{x_1}\sep \cdots
    \sep\enx[a_n]{}{}{x_n})\ (\psi^1 \land \ldots \land \psi^k)$
  \end{center}
  we have that $\mathcal{E}_1,C_1\models \psi$ while
  $\mathcal{E}_2,C_2\not\models \psi$. This gives the desired
  contradiction.  
  \qed
\end{proof}

Now it is immediate to conclude that the following holds.

\begin{theorem}[step bisimilarity]
  \label{th:step-bisimilarity}
  Let $\mathcal{E}_1$ and $\mathcal{E}_2$ be {\pes}s. Then
  $\mathcal{E}_1 \stepb \mathcal{E}_2$ iff $\mathcal{E}_1 \equiv_{\lcs}
  \mathcal{E}_2$.
\end{theorem}

\subsection{Pomset logic}
The logic $\lcp$ for pomset bisimilarity in Fig.~\ref{fig:Fragments}
consists of the fragment of $\lc$ where, still an event must be
immediately executed when quantified, but it is possible to refer to
dependencies between events. However, propositional connectives
(negation and conjunction) can be used only on closed formulae.


Roughly speaking, in $\lcp$ closed subformulae characterise
the execution of pomsets. Hence, the requirement that the
propositional operators are used only on closed subformulae prevents
pomset transitions from being causally linked to the events in the
past. These ideas are formalised by the results below.

First observe that a closed formula in $\lcp$ has always the shape 
\begin{center}
  $\enx[a_1]{\vx_1}{\vy_1}{z_1} \ldots
   \enx[a_n]{\vx_n}{\vy_n}{z_n}\ \psi$
\end{center}
where, if we let $Z = \{ z_1, \ldots, z_n \}$,
then $\vx_i, \vy_i \subseteq Z$ for any $i \in \{1, \ldots,n\}$.
We next prove that the prefix $\enx[a_1]{\vx_1}{\vy_1}{z_1} \ldots
\enx[a_n]{\vx_n}{\vy_n}{z_n}$ intuitively corresponds to the execution
of a class of pomsets (not a single one, since the relation between
some events might be not specified).
More precisely, in the situation above let
$\pomsets{\enx[a_1]{\vx_1}{\vy_1}{z_1} \ldots
  \enx[a_n]{\vx_n}{\vy_n}{z_n}}$ denote the class of pomsets $(Z,
\leq, \lambda)$ such that $Z = \{ z_1, \ldots, z_n\}$ and for $i \in
\{ 1, \ldots, n\}$, $\lambda(z_1) = \act_i$ and given any $z \in Z$
\begin{itemize}
\item $z \in \vx_i$ implies $z \leq z_i$,
\item $z \in \vy_i$ implies $z \not\leq z_i$.
\end{itemize}
With this definition it is immediate to show that the following result holds.

\begin{lemma}
  \label{le:weak-pomset}
  Let $\varphi = \enx[a_1]{\vx_1}{\vy_1}{z_1} \ldots
  \enx[a_n]{\vx_n}{\vy_n}{z_n}\ \psi$ be a closed formula in $\lcp$.
  Then 
  \begin{center}
     $\mathcal{E},C \models_\eta \varphi$
     \quad
     iff
     \quad
     \begin{tabular}[t]{l}
       $C \trans{X} C'$ where
       $X = \{ e_1, \ldots, e_n \}$ is a pomset s.t.
       $X \sim (Z, \leq, \lambda)$\\
       for some $(Z, \leq, \lambda) \in \pomsets{\enx[a_1]{\vx_1}{\vy_1}{z_1}
         \ldots \enx[a_n]{\vx_n}{\vy_n}{z_n}}$\\ 
       and
       $\mathcal{E},C' \models_{\eta'} \psi$,  
       with $\eta' = \eta[z_1 \mapsto e_1, \ldots, z_n \mapsto e_n]$
     \end{tabular}
  \end{center}
\end{lemma}

\begin{proof}
  By induction on $n$.
  \qed
\end{proof}

Next we observe that, in particular, the execution of a single pomset can be
exactly characterised by a corresponding formula in $\lcp$.

\begin{definition}[pomsets as formulae in $\lcp$]
  Let $Z = \{z_1,\ldots, z_n\}$ be a set of variables and let $\pom_Z
  = (Z,\leq_{\pom_Z},\lambda_{\pom_Z})$ be a pomset. Given a formula
  $\varphi \in \lcp$, we denote by $\enx[]{}{}{\pom_Z} \varphi$ the
  formula inductively defined as follows. If $Z$ is empty then
  $\enx[]{}{}{\pom_Z} \varphi = \varphi$. If $Z = Z' \cup \{ z \}$,
  where $z$ is maximal with respect to $\leq_{\pom_Z}$ (if there are
  many maximal $z_i$, choose the one with highest index), let $\vx =
  \{ z' \in Z' \mid z' \leq_{\pom_z} z \}$, $\vy = Z' \setminus \vx$,
  and $\act=\lambda_{\pom_Z}(z)$, then $\enx[]{}{}{\pom_Z} \varphi =
  \enx[]{}{}{\pom_{Z'}}\, \enx{\vx}{\vy}{z} \varphi$.
\end{definition}
Note that if $\varphi$ is a closed formula also $\enx[]{}{}{\pom_Z}
\varphi$ is closed.

The fact that pomset formulae as defined above have exactly
the intended semantics immediately follows from Lemma~\ref{le:weak-pomset}.

\begin{lemma}[pomsets in $\lcp$]
  \label{le:pomset}
  Let $\mathcal{E}$ be a {\pes} and let $C \in \conf{\mathcal{E}}$ be
  a configuration. Given $\{z_1,\ldots,z_n\} \subseteq \Var$ and a pomset
  $\pom_Z = (Z ,\leq_{\pom_Z},\lambda_{\pom_Z})$, then
  \begin{center}
     $\mathcal{E},C \models_\eta 
     \enx[]{}{}{\pom_Z}\, \varphi$
     \quad
     iff
     \quad
     \begin{tabular}[t]{l}
       $C \trans{X} C'$ where
       $X = \{ e_1, \ldots, e_n \}$ is a pomset s.t.
       $X \sim \pom_Z$\\ 
       and $\mathcal{E},C' \models_{\eta'} \varphi$,  
       with $\eta' = \eta[z_1 \mapsto e_1, \ldots, z_n \mapsto e_n]$
     \end{tabular}
  \end{center}
\end{lemma}

\begin{proof}
  Just observe that $\pomsets{\enx[]{}{}{\pom_Z}}=\{ \pom_Z \}$. Then the
  result is an instance of Lemma~\ref{le:weak-pomset}.
  \qed
\end{proof}


\begin{lemma}
  \label{le:pomset-logic}
  Let $\mathcal{E}_1$ and $\mathcal{E}_2$ be {\pes}s and let $C_i \in
  \conf{\mathcal{E}_i}$, for $i \in \{ 1,2\}$, be configurations.
  There exists a pomset bisimulation $R$ such that $(C_1, C_2) \in R$
  iff for any $\varphi \in \lcp$, $\varphi$ closed formula, $\mathcal{E}_1,
  C_1 \models \varphi \Leftrightarrow \mathcal{E}_2, C_2 \models
  \varphi$.
\end{lemma}

\begin{proof}
  ($\Rightarrow$) Let $R$ be a pomset bisimulation. We prove that if
  $(C_1,C_2) \in R$, then for all closed formulae $\varphi \in \lcp$,
  we have that $\mathcal{E}_1,C_1 \models \varphi$ iff
  $\mathcal{E}_2,C_2\models \varphi$.

  The proof proceeds by induction on the structure of the formula
  $\varphi$. The cases in which $\varphi$ is a conjunction, negation or
  true are trivial. In the remaining cases $\varphi$ is a closed
  formula of the shape
  \begin{equation}
    \label{eq:pomset}
    \enx[a_1]{\vx_1}{\vy_1}{z_1} \ldots
                          \enx[a_n]{\vx_n}{\vy_n}{z_n}\ \psi. 
  \end{equation}
  where $\psi$ is closed.
  %

  Assume that $\mathcal{E}_1,C_1 \models \varphi$, i.e.,
  $\mathcal{E}_1,C_1 \models_{\eta_1} \varphi$ for some (irrelevant)
  $\eta$.  Then, by Lemma~\ref{le:weak-pomset}, $C_1 \trans{X} C_1'$
  where $X \sim (Z, \leq, \lambda)$ for some pomset $ (Z, \leq,
  \lambda) \in \pomsets{\enx[a_1]{\vx_1}{\vy_1}{z_1} \ldots
    \enx[a_n]{\vx_n}{\vy_n}{z_n}}$.  Additionally $\mathcal{E}_1,C_1'
  \models_{\eta_1[z_1 \mapsto e_1, \ldots, z_n \mapsto e_n]} \psi$,
  which can be written $\mathcal{E}_1,C_1' \models \psi$, as $\psi$ is
  closed.

  Since $(C_1, C_2) \in R$ and $R$ is a pomset bisimulation, there is
  a pomset $Y = \{ g_1, \ldots, g_n\}$, isomorphic to $X$, and thus to
  $(Z,\leq,\lambda)$, such that
  \begin{equation}
    \label{eq:pomset1}
    C_2 \trans{Y} C_2'
  \end{equation}
  and $(C_1', C_2') \in R$.  By inductive hypothesis,
  $\mathcal{E}_2,C_2' \models \psi$. Again, since $\psi$ is closed, by
  Lemma~\ref{le:env-fv} it also holds $\mathcal{E}_2, C_2'
  \models_{\eta_2[z_1 \mapsto g_1, \ldots, z_n \mapsto g_n]} \psi$,
  for any chosen $\eta_2$. This fact, together with
  (\ref{eq:pomset1}), allows us to conclude, by
  Lemma~\ref{le:weak-pomset}, that $\mathcal{E}_2,C_2 \models_{\eta_2}
  \varphi$, i.e., since $\varphi$ is closed, $\mathcal{E}_2,C_2 \models
  \varphi$ as desired.

  \bigskip

  ($\Leftarrow$) The proof is analogous to that of
  Lemma~\ref{le:step-logic}, i.e., we show that the relation 
  \begin{center}
    $R = \{(C_1,C_2) \mid \forall \varphi \in \lcp, \mbox{$\varphi$
      closed},\ \mathcal{E}_1,C_1 \models \varphi$ iff
    $\mathcal{E}_2,C_2\models \varphi\}$
  \end{center}
  is a pomset bisimulation.

  We proceed by contradiction. Let $(C_1,C_2) \in R$, let
  $C_1\trans{X} C_1'$, where $X$ is a pomset, and assume that for all
  $Y$ such that $C_2\trans{Y}C_2'$ and $X \sim Y$ there exists a
  closed formula $\psi \in \lcp$ such that $\mathcal{E}_1,C_1' \models
  \psi$ and $\mathcal{E}_2,C_2' \not\models \psi$.
  
  Since our {\pes}s are assumed to be image finite, there are finitely
  many such pomset transitions $C_2 \trans{Y^i} C_2^i$, for $i \in
  \{1, \ldots, k\}$. Let $\psi^i$ be the formulae such that
  $\mathcal{E}_1,C'_1 \models \psi^i$ and $\mathcal{E}_2,C_2^i
  \not\models \psi^i$ for $i \in \{1, \ldots, k\}$.  If $\pom_Z$ is a
  pomset of variables, such that $\pom_Z \sim X$, let us define a
  formula in $\lcs$ as follows:
  \begin{center}
    $\psi = \enx[]{}{}{\pom_Z}\ (\psi^1 \land \ldots \land \psi^k)$
  \end{center}
  Then by Lemma~\ref{le:pomset}, we have that
  $\mathcal{E}_1,C_1\models \psi$ while $\mathcal{E}_2,C_2\not\models
  \psi$. This gives the desired contradiction.
  \qed
\end{proof}

The logical characterisation of pomset bisimilarity now immediately
follows.

\begin{theorem}[pomset bisimilarity]
  \label{th:pomset-bisimilarity}
  Let $\mathcal{E}_1$ and $\mathcal{E}_2$ be {\pes}s. Then
  $\mathcal{E}_1 \pomb \mathcal{E}_2$ iff $\mathcal{E}_1 \equiv_{\lcp}
  \mathcal{E}_2$.
\end{theorem}

As an example, consider the two {\pes}s $\mathcal{E}_7$ and
$\mathcal{E}_9$ in Fig.~\ref{fi:example2}.  They are step bisimilar
but not pomset bisimilar since only the second one can execute the
pomset $p_{a<b} = (\{\act, \act[b]\}, \act < \act[b], \lambda)$, where
$\lambda$ is the obvious labelling. Accordingly, the formula $\varphi
= \enx[]{}{}{p_{a<b}} \true = \enx{}{}{x}\enx[b]{x}{}{y} \true$ in
$\lcp$, is satisfied only by $\mathcal{E}_9$.

\subsection{History preserving logic}

The fragment $\lchp$ corresponding to hp-bisimilarity is essentially
the same as for pomset logic, where we relax the condition asking that
the propositional connectives are applied only to closed
formulae. Intuitively, in this way a formula $\varphi\in\lchp$,
besides expressing the possibility of executing a pomset $\pom$, also
predicates about dependencies of events in the pomset with previously
executed events (bound to the free variables of $\varphi$).
%

The following two
  {\pes}s can be proved to be pomset equivalent but not hp-equivalent:
  \begin{center}
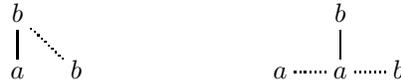
\begin{tabular}{cc}
\mbox{
\xymatrix@R=3mm@C=4mm{
 b  &      \\
a\ar@{-}[u]  & b\ar@{.}[ul] \\
}}
&\hspace*{2cm}
\mbox{
\xymatrix@R=3mm@C=4mm{
    &   b      &     \\
a \ar@{.}[r]   &   a \ar@{-}[u]      &  b\ar@{.}[l]
}}
\end{tabular}
\end{center}

Intuitively, they allow the same pomset transitions, but they have a
different ``causal branching''. Indeed, only in the left-most {\pes}s,
after the execution of an $\act$-labelled event we can choose between
an independent and a dependent $\act[b]$-labelled event. In the
rightmost {\pes} the choice is already determined by the execution of
$\act$. Formally, the formula $\enx{}{}{x}(\enx[b]{}{x}{y}\true \wedge
\enx[b]{x}{}{z} \true)$ in $\lchp$ is true only on the left-most
{\pes}.


We start with a lemma that makes explicit the semantics of the induced
operator $\enx{\vx}{\vy}{z}$.

\begin{lemma}[events with their history in the logic]
  \label{lem:HistToLogic}
  Given a {\pes} $\mathcal{E}$, a formula $\varphi\in\lchp$ and a
  legal pair $(C,\eta) \in \ld{\enx{\vx}{\vy}{z}\, \varphi}$:
  \begin{center}
    $\mathcal{E},C \models_\eta 
    \enx{\vx}{\vy}{z}\, \varphi$
    \quad
    iff
    \quad
    \begin{tabular}{l}
      there is an event $e \in E$ such that 
      $C \trans{e} C'$, $\lambda(e)=\act$,\\
      $\eta(\vx)\leq e$, $\eta(\vy)\conc e$ and
      $C' \models_{\eta'} \varphi$,  where $\eta' = \eta[z \mapsto e]$.
    \end{tabular}
  \end{center}
\end{lemma}

\begin{proof}
  The result follows almost immediately from the definition of the
  semantics (Definition~\ref{de:semantics}).
  \qed
\end{proof}

\begin{lemma}
  \label{le:hp-logic}
  Let $\mathcal{E}_1$ and $\mathcal{E}_2$ be {\pes}s and let $(C_1, f,
  C_2) \in \conf{\mathcal{E}_1} \bar{\times} \conf{\mathcal{E}_2}$,
  i.e., $C_i \in \conf{\mathcal{E}_i}$, for $i \in \{ 1, 2\}$, are
  configurations and $f : C_1 \to C_2$ is an isomorphism of
  pomsets.  Then the following are equivalent:
  \begin{enumerate}
  \item there is a hp-bisimulation $R$ such that $(C_1, f, C_2) \in R$;
  \item for any $\varphi \in \lchp$ and $\eta_1 \in
    \Env_{\mathcal{E}_1}$ such that $\eta_1(\fv{\varphi}) \subseteq
    C_1$, it holds that $\mathcal{E}_1, C_1 \models_{\eta_1} \varphi
    \Leftrightarrow \mathcal{E}_2, C_2 \models_{f \circ \eta_1}
    \varphi$.
  \end{enumerate}
\end{lemma}

\begin{proof}
  (1 $\Rightarrow$ 2) Let $R$ be a hp-bisimulation. We show that for
  all formulae $\varphi\in\lchp$, triples $(C_1,f,C_2) \in R$ and
  environments $\eta_1 \in \Env_{\mathcal{E}_1}$ such that
  $\eta_1(\fv{\varphi}) \subseteq C_1$ it holds
  \begin{center}
    $\mathcal{E}_1,C_1 \models_{\eta_1} \varphi\ $ iff
    $\ \mathcal{E}_2,C_2\models_{f \circ \eta_1} \varphi$.
  \end{center}
  We proceed by induction on the structure of the formula
  $\varphi$. We focus on the only non-trivial case where
  $\varphi=\enx{\vx}{\vy}{z}\, \psi$. If $\mathcal{E}_1,C_1
  \models_{\eta_1} \varphi$, then by Lemma~\ref{lem:HistToLogic} there is
  an event $e \in E_1$ such that 
  \begin{equation}
    \label{eq:hp1}
    C_1\trans{e} C_1'
  \end{equation}
  with $\lambda_1(e)=\act$, $\eta_1(\vx)\leq e$, $\eta_1(\vy)\conc
  e$ and $\mathcal{E}_1,C'_1 \models_{\eta_1'} \psi$ 
  where $\eta_1'=\eta_1[z\mapsto e]$.

  Since $(C_1, f, C_2) \in R$, there exists an event $g \in E_2$ such that
    \begin{equation}
    \label{eq:hp2}
    C_2\trans{g}C'_2
  \end{equation}
  and $(C_1',f',C_2')\in R$, with $f'=f[e\mapsto g]$.  Since $f'$ is an isomorphism of configurations,
  we have that $\lambda_2(g)=\act$, $f(\eta_1(\vx))\leq g$ and
  $f(\eta_1(\vy))\conc g$.

  Note that $\eta_1'(\fv{\psi}) \subseteq \eta_1'(\fv{\varphi} \cup \{ z
  \}) = \eta_1(\fv{\varphi}) \cup \{e\} \subseteq C_1 \cup \{ e \} =
  C_1'$. Thus, we can use the induction hypothesis to deduce that
  $\mathcal{E}_2,C'_2 \models_{f'\circ\eta_1'} \psi$.
  Therefore, by using again Lemma~\ref{lem:HistToLogic}, we can
  conclude $\mathcal{E}_2,C_2\models_{f \circ \eta_1} \varphi$.

  \smallskip

  The proof that $\mathcal{E}_2, C_2\models_{f \circ \eta_1} \varphi$
  implies $\mathcal{E}_1,C_1 \models_{\eta_1} \varphi$ is analogous and
  thus omitted.

  \bigskip
  \noindent
  (1 $\Leftarrow$ 2) As in Proposition~\ref{pr:LogicToHhp} we fix a
  surjective environment $\eta_1 : \Var \to E_1$. Moreover, given an
  event $e \in E_1$, we write $x_{e}$ to denote a fixed
  distinguished variable such that $\eta_1(x_{e}) = e$. Similarly,
  for a configuration $C_1=\{e_1,\ldots,e_n\}$ we denote by $X_{C_1}$
  the set of variables $\{ x_{e_1}, \ldots, x_{e_n}\}$. Observe that
  $(C_1,\eta_1)$ is a legal pair for any formula $\varphi\in\lc$ such
  that $\fv{\varphi}\subseteq X_{C_1}$.
  
  Then we show that the posetal relation $R \subseteq
  \conf{\mathcal{E}_1} \bar{\times} \conf{\mathcal{E}_2}$ defined by
  \begin{center}
    $R =\{(C_1,f,C_2) \mid \forall \varphi\in\lchp.\
    \fv{\varphi}\subseteq X_{C_1}\ \ \ \mathcal{E}_1,C_1
    \models_{\eta_1} \varphi$ iff
    $\mathcal{E}_2,C_2\models_{f \circ \eta_1} \varphi\}$
  \end{center}
  is a hp-bisimulation.
  Note that as in Proposition~\ref{pr:LogicToHhp}, with a slight abuse
  of notation, we denote by $f \circ \eta_1$ any environment $\eta_2$
  such that $\eta_2(x) = f (\eta_1(x))$ for $x \in X_{C_1}$ and
  $\eta_2(x)$ has any value, otherwise.
  By Lemma~\ref{le:env-fv}, this arbitrariness has no impact on the
  satisfaction of $\varphi$ in the definition of $R$ since
  $\fv{\varphi} \subseteq X_{C_1}$.

  We proceed by contradiction. Assume that $(C_1,f,C_2)\in R$, let
  $C_1\trans{e} C'_1$ and suppose that for all $g \in E_2$ such that
  $C_2\trans{g} C'_2$ with $C_1' \sim C_2'$ as pomsets, we have
  $(C_1', f[e \mapsto g], C_2') \not\in R$, i.e., there exists a
  formula $\psi$, with $\fv{\psi} \subseteq X_{C_1'}$, such that
  $\mathcal{E}_1, C_1' \models_{\eta_1} \psi$ and $\mathcal{E}_2,C_2'
  \not\models_{f' \circ \eta_1} \psi$. 
  
  Since all {\pes}s are assumed to be image finite, there are
  finitely many transitions
  \begin{center}
    $C_2\trans{g^i}C_2^i$, \quad $i \in \{ 1, \ldots, k\}$
  \end{center}
  such that $f^i = f[e \mapsto g^i]: C_1' \to C_2^i$ is an
  isomorphism of pomsets. Let $\psi^i$, for $i \in \{1,\ldots,k\}$ be
  formulae such that
  \begin{center}
    $\mathcal{E}_1,C_1' \models_{\eta_1} \psi^i$ \quad and \quad
    $\mathcal{E}_2,C^i_2 \not\models_{f^i\circ\eta_1} \psi^i$
  \end{center}
  where $\fv{\psi^i}\subseteq X_{C_1'}=X_{C_1}\cup\{x_{e}\}$.
  Now consider the formula
  $$
  \varphi=\enx{\vx}{\vy}{x_{e}}(\psi^1\land\ldots\land\psi^k)
  $$ 
  where $\act=\lambda_1(e)$ and the $\vx,\vy\subseteq X_{C_1}$
  are such that $\eta_1(\vx)$ is the set of causes of $e$ in
  $C_1$ and $\eta_1(\vy)$ is the set of events in $C_1$ which are
  concurrent with $e$.  
  Note that $\fv{\varphi} = \vx \cup \vy \cup (( \bigcup_{i=1}^k
  \fv{\psi_i}) \setminus \{x_{e}\}) = X_{C_1}$.

  Then by Lemma~\ref{lem:HistToLogic} we have
  that $\mathcal{E}_1,C_1\models_{\eta_1}\varphi$ and
  $\mathcal{E}_2,C_2\not\models_{f \circ \eta_1}\varphi$,
  which gives the desired contradiction.
  
  \medskip

  The fact that $R$ as defined above is a hp-bisimulation allows us to
  conclude. In fact, assume that $(C_1, f, C_2) \in
  \conf{\mathcal{E}_1} \bar{\times} \conf{\mathcal{E}_2}$ and (2)
  holds. Then for any $\varphi \in \lchp$ such that
  $\fv{\varphi}\subseteq X_{C_1}$, it holds that $\eta_1(\fv{\varphi})
  \subseteq \eta_1(X_{C_1}) = C_1$. Therefore we can use (2) and
  deduce that $\mathcal{E}_1,C_1 \models_{\eta_1} \varphi$ iff
  $\mathcal{E}_2,C_2 \models_{f \circ \eta_1} \varphi$. This
  implies that $(C_1, f, C_2) \in R$, i.e., we get (1).
  \qed
\end{proof}

\paragraph{Remark.}{It is worth observing that the hp-bisimulation
  built in the previous proof relates two configurations $C_1$ and
  $C_2$ when they satisfy the same formulae, whereas the
  hhp-bisimulation built in the proof of
  Proposition~\ref{pr:LogicToHhp} (which leads to
  Theorem~\ref{th:hhp-bisimilarity}) relates $C_1$ and $C_2$ when the
  same formulae are satisfied by the empty configuration (in an
  environment that binds free variables to $C_1$, resp. $C_2$).
  Intuitively, this corresponds to the fact that for hp-bisimilarity
  one has to check only the future of a configuration, while for
  hhp-bisimilarity also alternative evolutions (hence evolutions from
  the past) of a configuration must be considered.  }

\begin{theorem}[hp-bisimilarity]
  \label{th:hp-bisimilarity}
  Let $\mathcal{E}_1$ and $\mathcal{E}_2$ be {\pes}s. Then
  $\mathcal{E}_1 \hpb \mathcal{E}_2$ iff $\mathcal{E}_1 \equiv_{\lchp}
  \mathcal{E}_2$.
\end{theorem}

\begin{proof}
  ($\Rightarrow$)
  Let $\mathcal{E}_1 \hpb \mathcal{E}_2$. Then there is a
  hp-bisimulation $R$ such that $(\emptyset, \emptyset, \emptyset) \in
  R$. For all $\varphi \in \lchp$, if $\varphi$ is closed, i.e.,
  $\fv{\varphi} = \emptyset$, as an instance of
  Lemma~\ref{le:hp-logic}, we obtain $\mathcal{E}_1, \emptyset
  \models_{\eta_1} \varphi$ iff $\mathcal{E}_2, \emptyset \models_{f
    \circ \eta} \varphi$, for any $\eta_1 \in
  \Env_{\mathcal{E}_1}$. This amounts to $\mathcal{E}_1 \models
  \varphi$ iff $\mathcal{E}_2 \models \varphi$, i.e., $\mathcal{E}_1
  \equiv_{\lchp} \mathcal{E}_2$, as desired.

  ($\Leftarrow$)
  Let $\mathcal{E}_1 \equiv_{\lchp} \mathcal{E}_2$. Then, for any
  closed formula $\varphi \in \lchp$, it holds that $\mathcal{E}_1
  \models \varphi$ iff $\mathcal{E}_2 \models \varphi$. Since
  $\varphi$ is closed, satisfaction does not depend on the
  environment, hence $\mathcal{E}_1, \emptyset \models_{\eta_1}
  \varphi$ iff $\mathcal{E}_2, \emptyset \models_{\eta_2} \varphi$ for
  any $\eta_1 \in \Env_{\mathcal{E}_1}$, $\eta_2 \in
  \Env_{\mathcal{E}_2}$. In particular, we can consider $\emptyset :
  \emptyset \to \emptyset$, isomorphism between empty configurations
  and we have $\mathcal{E}_1, \emptyset \models_{\eta_1} \varphi$ iff
  $\mathcal{E}_2, \emptyset \models_{\emptyset \circ \eta_1} \varphi$
  for any $\eta_1 \in \Env_{\mathcal{E}_1}$. Therefore, we can apply
  Lemma~\ref{le:hp-logic} to conclude that there exists a
  hp-bisimulation $R$ such that $(\emptyset, \emptyset, \emptyset) \in
  R$ and thus $\mathcal{E}_1 \hpb \mathcal{E}_2$.
\qed
\end{proof}


\section{A logic with recursion: $\lcmu$}
\label{se:recursion}

The logic $\lc$ discussed in the previous section is theoretically
interesting as it allows one to logically characterise the main true
concurrent equivalences. However, as a specification language, it has
a limited expressiveness: even if one can ``observe'' events
arbitrarily far in the future, a single formula in $\lc$ only 
describes properties where a finite number of events are
executed.
In order to overcome this limitation, in this section we study a
fixpoint extension of the logic, where the use of recursion allows one
to express causal and concurrency properties of infinite
computations. The resulting logic, denoted $\lcmu$, is a kind of
first-order $\mu$-calculus similar to the $\mu$-calculi
in~\cite{Dam:MCMP,DFG:TPVODS} and~\cite{GW:MCPD}, where first order
variables are used to represent channels or data. Similarities exist
also with the fixpoint extension of independence-friendly modal logic
studied in~\cite{BK:CIFFL}. In fact, in all of these papers fixpoints
are added to a core logic which includes quantified first order
variables. The solutions adopted to let the fixpoint operators and
variables interact with first order variables is similar to that in
our logic. 

Let $\aProp$ be a set of \emph{abstract propositions}, ranged over by
$X$, $Y$, \ldots, that are intended to represent formulae possibly
containing (unnamed) free event variables. Each abstract proposition
has an arity $\art{X}$, which indicates the number of free event
variables in $X$. An abstract proposition $X$ can be turn into a
formula by specifying a name for its free variables. For $\vx$ such
that $|\vx| = \art{X}$, we write $\p{X}{\vx}$ to indicate the abstract
proposition $X$ whose free event variables are named $\vx$. We call
$\p{X}{\vx}$ a \emph{proposition} and denote by $\Prop$ the set of all
propositions. 



\begin{definition}[syntax]
  \label{de:syntax-mu}
  Let $\Var$ be a denumerable set of event variables and let $\Prop$
  be a set of propositions, as explained above.  The syntax of $\lcmu$
  over the set of labels $\Lambda$ is defined as follows:
  $$
  \begin{array}{ll}
    \varphi\ ::=\  
    \p{X}{\vx} \  \mid\ &  
    \true\ \mid\ \varphi \land \varphi\ \mid\ \neg\varphi\ \mid\ 
    \en{\vx}{\vy}{z}\, \varphi\  \mid\ 
    \ex{z}\varphi \mid\ \mu \p{X}{\vx}. \varphi\
  \end{array}
  $$ 
  where for formula $\mu
  \p{X}{\vx}. \varphi$, as usual, $X$ must occur positively
  in $\varphi$ and additionally, $\fv{\varphi} = \vx$.
\end{definition}

The requirement that $X$ occurs positively in the formula $\mu
\p{X}{\vx}. \varphi$ is a standard one, later used in the definition of the
semantics for ensuring the existence of the fixpoint.

\begin{definition}[free variables]
  The \emph{free variables} of a formula $\varphi$ in $\lcmu$ are
  given as in Definition~\ref{de:free-vars}, with the addition of the
  following clauses:
  \begin{center}
    $\fv{\p{X}{\vx}} = \vx$ 
    \qquad and 
    \qquad 
    $\fv{\mu \p{X}{\vx}.\varphi} = 
  \vx$.
  \end{center}
\end{definition}
%
In the following we will often use the set of free variables of a
formula as a tuple. Thus it is convenient to assume that $\fv{\cdot}$ returns a fixed tuple of variables.
Note that the fact that variables $\vx$ are free in $\p{X}{\vx}$ and
in $\mu \p{X}{\vx}.\varphi$ is reflected in the definition of free
variable substitution. For instance $\p{X}{\vx}[\vy/\vx] = \p{X}{\vy}$
and $(\mu \p{X}{x}.\varphi)[y/x]=\mu \p{X}{y}.(\varphi[y/x])$.

A least fixpoint operator $\mu$ has been added. In a recursive formula
$\mu \p{X}{\vx}. \varphi$ the abstract proposition $X$ can occur in
$\varphi$, possibly with a different tuple of variables which,
intuitively, are used in the next iteration.

As usual a greatest fixpoint operator can be encoded, by duality, as
\begin{center}
  $\nu \p{X}{\vx}. \varphi = \neg (\mu \p{X}{\vx}. \neg \tilde{\varphi})$
\end{center}
where $\tilde{\varphi}$ is the formula obtained replacing any
occurrence of $X$ in $\varphi$ with $\neg X$ (in
order to keep the positivity of the occurrences of $X$).


As an example, the existence of a run consisting of an infinite causal
chain of $\act$-actions can be expressed by the following formula:
$$
\enx{}{}{x} \ (\nu X(x). \enx{x}{}{y} \p{X}{y})
$$ 
The infinite causal chain is obtained by ``passing'' the event
bound to $y$ by the current execution to the next iteration so
that it can be used as a cause in the corresponding execution.
The execution outside the recursive formula binds $x$ to an
$\act$-labelled event which will be the first in the causal chain.

In a fixpoint formula $\mu \p{X}{\vx}. \varphi$, the fixpoint
operator binds all the free occurrences of the abstract proposition
$X$ in $\varphi$. This leads to the following notion of free abstract
proposition.

\begin{definition}[free propositions, substitution]
  The set of \emph{free propositions} in a formula $\varphi$ in
  $\lcmu$, denoted $\fp{\varphi}$, is defined inductively by
  \begin{quote}
    $\fp{\true}   =  \emptyset$ \qquad $\fp{\p{X}{\vx}}  =  \{ X \}$\\[1mm]
    $\fp{\varphi_1\land\varphi_2} =  \fp{\varphi_1} \cup \fp{\varphi_2}$\\[1mm]
    $\fp{\neg \varphi} = 
    \fp{\en{\vx}{\vy}{z}\, \varphi}  = 
    \fp{\ex{z}\varphi}  =  \fp{\varphi}$\\[1mm]
    $\fp{\mu \p{X}{\vx}. \varphi} =  
    \fp{\varphi} \setminus \{X \}$
  \end{quote}
  Let $\varphi$ be a formula in $\lcmu$. For an abstract proposition
  $X$ and formula $\psi$ such that $\fv{\psi} =\vx$, $|\vx| =
  \art{X}$, we denote by $\varphi[\psi/X]$ the formula obtained from
  $\varphi$ by replacing any free occurrence of $\p{X}{\vy}$ by
  $\psi[\vy/\vx]$.
\end{definition}

 A formula $\varphi \in \lcmu$ is called \emph{closed} when both $\fv{\varphi}$
 and $\fp{\varphi}$ are empty.

\medskip 
Let us now move to the definition of the semantics. Legal pairs for a
formula are defined exactly as in Definition~\ref{de:legal-pairs}. For
instance the pair $(C,\eta)$ is legal for the formula $\p{X}{\vx}$
if the set $C\cup\eta(\vx)$ is consistent. 
On the other hand, in addition to the (event variable) environment,
the semantics of $\lcmu$ also requires an interpretation for the
propositions, mapping each proposition $\p{X}{\vx}$ to a set of legal
pairs for it.
%

\begin{definition}[proposition environments]
  \label{def:propEnv}
  Let $\mathcal{E}$ be a {\pes}.
  A \emph{proposition environment} is a
  function 
  $\pi : \mathcal{X} \to 2^{\conf{\mathcal{E}} \times \Env_{\mathcal{E}}}$
  such that:
  \begin{enumerate}
  \item  
    $\pi(\p{X}{\vx}) \subseteq \ld{\p{X}{\vx}}$ 
    for any $\p{X}{\vx}\in\mathcal{X}$, and
    
  \item if $(C, \eta) \in \pi(\p{X}{\vx})$ and $\eta'(\vy) = \eta(\vx)$ pointwise,
    then $(C, \eta') \in \pi(\p{X}{\vy})$.
  \end{enumerate}
  We denote by $\PEnv_\mathcal{E}$ the set of proposition environments, ranged
  over by $\pi$.
\end{definition}
The first condition requires that the denotation for $\p{X}{\vx}$ only
consists of legal pairs for $\p{X}{\vx}$. The second condition
requires that the semantics of a proposition only depends on the
events that the environment associates to its free variables and that
it does not depend on the naming of the variables.  Such a condition
allows us to generalise Lemma~\ref{le:env-fv} to the logic with recursion.

Updates of a proposition environment must be properly defined in order to
maintain the validity of properties 1 and 2 above. For $\pi \in
\PEnv_{\mathcal{E}}$ and $S \subseteq \ld{\p{X}{\vx}}$, we write
$\pi[\p{X}{\vx} \mapsto S]$ for the proposition environment defined by
\begin{quote}
  \begin{tabular}{lll}
    $\pi[\p{X}{\vx} \mapsto S] (\p{X}{\vy})$ & = & $\{ (C, \eta') \mid (C, \eta) \in S\ \land\  \eta'(\vy) = \eta(\vx) \}$\\[1mm]
  $\pi[\p{X}{\vx} \mapsto S] (\p{Y}{\vy})$ & = & $\pi(\p{Y}{\vy})$  \quad for $Y \neq
X$.
  \end{tabular}
\end{quote}

\begin{lemma}
  \label{le:subst-mu}
  Let $\mathcal{E}$ be a {\pes}, $\pi$ a proposition environments,
  $\varphi \in \lcmu$ be a formula and let $\vx = \fv{\varphi}$ be the
  tuple of free variables in $\varphi$.
  
  \begin{enumerate}
  \item 
    If $(C, \eta) \in \sem[\mathcal{E}]{\varphi}_\pi$ and $\eta'(\vy) =
    \eta(\vx)$ pointwise, then $(C, \eta') \in
    \sem[\mathcal{E}]{\varphi[\vy/\vx]}_\pi$.
    
  \item For any formula $\psi$ and abstract proposition $X$ such that
    $\art{X} = |\fv{\varphi}|$ it holds
    $\sem[\mathcal{E}]{\psi[\varphi/X]}_\pi =
    \sem[\mathcal{E}]{\psi}_{\pi[\p{X}{\vx} \mapsto
      \sem[\mathcal{E}]{\varphi}_\pi]}$.
  \end{enumerate}
\end{lemma}

\begin{proof}
  Both items can be proved by a routine induction (on $\varphi$ for 1
  and on $\psi$ for 2).  \qed
\end{proof}

In particular, from 1 above it follows that, as already proved for
logic $\lc$ in Lemma~\ref{le:env-fv}, the semantics of a formula $\varphi$
in $\lcmu$ only depends on the events that the environment associates
to the free variables $\vx$ of the formula, i.e., if $C \in
\conf{\mathcal{E}}$ and $\eta, \eta'$ are environments such that
$\eta_{|\vx} = \eta'_{|\vx}$ then $(C, \eta) \in
\sem[\mathcal{E}]{\varphi}$ iff $(C, \eta') \in
\sem[\mathcal{E}]{\varphi}$.

\begin{definition}[semantics]
  \label{de:semantics-mu}
  Let $\mathcal{E}$ be a {\pes}. The
  denotation of a formula is given by the function
  \begin{center}
    $\sem[\mathcal{E}]{\cdot}: \lcmu \to \PEnv_{\mathcal{E}} \to 2^{\conf{\mathcal{E}} \times \Env_{\mathcal{E}}}$
  \end{center}
  defined inductively as follows, where we write
  $\sem[\mathcal{E}]{\varphi}_\pi$ instead of $\sem[\mathcal{E}]{\varphi}(\pi)$:
  \begin{center}
    $
    \begin{array}{rcl}
      \sem[\mathcal{E}]{\true}_\pi & = &  \conf{\mathcal{E}} \times
      \Env_{\mathcal{E}}
      \\[3mm]
      %
      %
      \sem[\mathcal{E}]{\varphi_1 \land \varphi_2}_\pi & = &  
      \sem[\mathcal{E}]{\varphi_1}_\pi \cap \sem[\mathcal{E}]{\varphi_2}_\pi \cap \ld{\varphi_1 \land \varphi_2}\\[3mm]
      %
      %
      \sem[\mathcal{E}]{\neg \varphi}_\pi & = &  
      \ld{\varphi} \setminus \sem[\mathcal{E}]{\varphi}_\pi\\[3mm]
      \sem[\mathcal{E}]{\en{\vx}{\vy}{z}\, \varphi}_\pi & = &
      \{ (C, \eta) \mid 
      \begin{array}[t]{ll}
        (C, \eta) \in \ld{\en{\vx}{\vy}{z}\, \varphi}\ \mathrm{and}\\
        \exists e \in \res{E}{C} \mathrm{\ such\ that\ }\ e \cons \eta(\fv{\varphi} \setminus \{ z \})\\
        \land\ \lambda(e) = \act\ \land\ \eta(\vx) < e\ \land\ \eta(\vy) \conc e\\
        \land\  (C, \eta[z\mapsto e]) \in
        \sem[\mathcal{E}]{\varphi}_\pi  & \}\\[3mm]
      \end{array}
      \\
      %
      %
      \sem[\mathcal{E}]{\ex{z}\, \varphi}_\pi & = &
      \{ (C, \eta) \mid 
      \begin{array}[t]{ll}
        C \trans{\eta(z)} C'\ \land\  (C', \eta) \in
        \sem[\mathcal{E}]{\varphi}_\pi & \}\\[2mm]
      \end{array}
      \\
      %
      %
      %
      \sem[\mathcal{E}]{\p{X}{\vx}}_\pi & = &  \pi(\p{X}{\vx})
      \\[3mm]
      \sem[\mathcal{E}]{\mu \p{X}{\vx}. \varphi}_\pi 
      & = &
      \mathit{lfp}(f)
    \end{array}
    $
  \end{center}
  \noindent
  where $\mathit{lfp}(f)$ is the least fixed point of the function $f
  : 2^{\ld{\p{X}{\vx}}} \to 2^{\ld{\p{X}{\vx}}}$ that maps 
   $S \subseteq \ld{\p{X}{\vx}}$ into   
  \begin{center}
    $f(S) =  \sem[\mathcal{E}]{\varphi}_{\pi[\p{X}{\vx} \mapsto S]}$
  \end{center}

  When $(C, \eta) \in \sem[\mathcal{E}]{\varphi}_\pi$ we say that the
  {\pes} $\mathcal{E}$ satisfies the formula $\varphi$ in the
  configuration $C$ and environments $\eta,\pi$ and write
  $\mathcal{E}, C \models_{\eta,\pi} \varphi$. For closed formulae
  $\varphi$, we write $\mathcal{E}, C \models \varphi$, when
  $\mathcal{E}, C \models_{\eta,\pi} \varphi$ for some $\eta$, $\pi$
  and $\mathcal{E} \models \varphi$ when $\mathcal{E}, \emptyset
  \models \varphi$.
\end{definition}

\noindent
It can be easily proved that Lemma~\ref{le:legal} extends to $\lcmu$,
i.e., for any formula $\varphi\in\lcmu$, its denotation only contains
legal pairs, that is $\sem[\mathcal{E}]{\varphi}_\pi \subseteq
\ld[\mathcal{E}]{\varphi}$.
Note also that the semantics of recursive formulae is well-defined. In fact,
$\pi[\p{X}{\vx} \mapsto S]$ is a well-defined
proposition environment, since $S \subseteq \ld{\p{X}{\vx}}$.
Moreover $f(S) = \sem[\mathcal{E}]{\varphi}_{\pi[\p{X}{\vx} \mapsto
    S]} \subseteq \ld{\varphi}$ by the previous observation, and
$\ld{\p{X}{\vx}}=\ld{\varphi}$ since $\fv{\varphi} = \vx$ by
definition of the syntax of $\lcmu$. Therefore, correctly, $f(S)
\subseteq \ld{\p{X}{\vx}}$.
Moreover, the least fixed point of $f$ exists by
Knaster-Tarski theorem since the set $2^{\ld{\p{X}{\vx}}}$ ordered by
subset inclusion is a complete lattice and the function $f$ used in
the definition is monotone. This can be easily checked by inspection
of the definition of the semantics (Definition~\ref{de:semantics-mu}),
keeping in mind that $X$ is required to occur positively in $\varphi$.

\medskip
As it happens for the non-recursive fragment $\lc$, the logic
$\lcmu$ could be defined in positive form. The corresponding syntax,
given below, includes the dual operators and omits negation, which can
then be encoded by duality.
\[
  \begin{array}{llllll}
    \varphi\ ::=\  
    \p{X}{\vx} \  \mid\ &  
    \true\ \mid\ 
    & \varphi \land \varphi\ \mid\ 
    & \en{\vx}{\vy}{z}\, \varphi\  \mid\ 
    & \ex{z}\varphi \mid\ 
    & \mu \p{X}{\vx}. \varphi\\[2mm]
    & \false\ \mid\  
    & \varphi \lor \varphi\ \mid\
    & \enBox{\vx}{\vy}{z}\, \varphi\ \mid\ 
    & \exBox{z}\varphi \mid\ 
    & \nu \p{X}{\vx}. \varphi
  \end{array}
\]
In the following we will
freely use the dual operators.

\subsection{Examples}

In the previous section we observed that standard HM logic can be
viewed as a fragment of $\lc$ where we only use the (derived) modality
$\enx{}{}{x}$. Similarly, the propositional $\mu$-calculus corresponds
to a fragment of the the general logic $\lcmu$ where we avoid references
to causally dependent/independent events. In particular, since in
recursive formulae we do not express causal links between event
variables used in different iterations, we can use only propositions
without free variables (i.e., of arity $0$). Therefore, the
$\mu$-calculus corresponds to the following fragment of $\lcmu$:
\begin{center}
    $\varphi\ ::=\  
    \p{X}{\epsilon} \  \mid\  
    \true\ \mid\ \varphi \land \varphi\ \mid\ \neg\varphi\ \mid\
    \en{\vx}{\vy}{z}\, \varphi\  \mid\ 
    \ex{z}\varphi \mid\ \mu X(\epsilon). \varphi$
\end{center}
For simplicity in the
following we omit trailing empty tuples of variables, writing $X$ instead of $\p{X}{\epsilon}$.

As first examples of $\lcmu$ formulae we thus have some standard
safety and liveness properties inherited from the $\mu$-calculus (see,
e.g.,~\cite{BS:MMC}). For a fixed closed formula $\psi$, representing a
property of interest:
\begin{itemize}
\item $\psi$ holds in every reachable state\\
  $\mathit{Inv}(\psi) = \nu X.\ ( \psi\land\enxBox[\_]{}{}{z}X)$;

\item $\psi$ eventually holds in some state\\
  $\mathit{Pos}(\psi) = \mu X.\ ( \psi\lor\enx[\_]{}{}{z}X)$;

\item there is a
  complete (finite terminated or infinite) computation where $\psi$ always
  holds\\
  $\mathit{Safe}(\psi) = \nu X.\ ( \psi\land
  (\enxBox[\_]{}{}{z}\false\lor\enx[\_]{}{}{x}X))$;

\item  in every complete computation eventually $\psi$ holds\\
  $\mathit{Ev}(\psi) =
  \mu X.\ ( \psi\lor
  (\enx[\_]{}{}{z}\true\land\enxBox[\_]{}{}{x}X))$. 
\end{itemize}

\smallskip

When moving to the full logic, property $\psi$ can include
concurrency and causal features. In case $\psi$ is not closed, denoted
by $\vx$ the tuple of free variables in $\psi$, in order to respect
the syntax any occurrence of $X$ above must be replaced by $X(\vx)$.
For instance, we can define
$\mathit{Ev}((\enx{}{}{z} \otimes \enx{}{}{z'}) \true)$ saying that eventually
there will be a concurrent step consisting of two events, labelled
$\act$ and $\act[b]$, respectively, or $ \mathit{Inv}(\enx[r]{}{}{z}
\mathit{Ev}(\enx[s]{z}{}{z'} \true))$ saying that any $\act[r]$-labelled event will be eventually followed by an $\act[s]$-labelled event caused by it (e.g., any request will be eventually served).

\smallskip

More generally, logic $\lcmu$ allows one to express causal and
concurrency properties of infinite computations, where events occurring
in different fixpoint iterations are possibly related. We next provide a
number of further examples.

\begin{itemize}
\item There is a causal chain of
  $\act[b]$-labelled events reaching a state where $\act$ can be fired:
   $$
   \enx{}{}{y}\true \lor \, \enx[b]{}{}{x}\, 
   (\mu X(x).(\enx{}{}{z}\true \lor\enx[b]{x}{}{y} \, \p{X}{y}))
   $$

 \item There is an executable $\act$-labelled event
   such that in every configuration reached by executing events which are
   concurrent with it, a $\act[c]$-labelled event can be executed:
   $$
   \en{}{}{x} (\ex{x}\true \land \nu X(x).(\enx[c]{}{}{z} \true 
   \land \enxBox[\_]{}{x}{y}\, \p{X}{x}))
   $$

 \item
   It is always possible to perform
   a step consisting of two concurrent events labelled by $\act$ and
   $\act[b]$, after executing any number of events labelled $\act[c]$: 
   $$
   \nu X.\ ((\enx[a]{}{}{z} \sep \enx[b]{}{}{z'}) \true  \land \enxBox[c]{}{}{w} X)
   $$

 \item There is a finite sequence of (not necessarily related) steps,
   each consisting of two concurrent events labelled by $\act$ and
   $\act[b]$, respectively, leading to a state where a
   $\act[c]$-labelled event can be executed:
   \[
   \mu X. (\enx[c]{}{}{z}\true  \lor (\enx[a]{}{}{z} \sep \enx[b]{}{}{z'}) \, X)
   \]
   
\end{itemize}

\subsection{Invariance of logical equivalence}
\label{ss:invariance}

We show that the addition of fixpoints formulae does not alter the
logical equivalence, that still coincides with hhp-bisimilarity, i.e.,
$\equiv_{\lc}\, =\, \equiv_{\lcmu}\, =\, \hhpb$.  (Recall that in the
paper we are limiting ourselves to image-finite $\pes$s.)  This is
done by adapting the proof of the fact that the $\mu$-calculus induces
the same equivalence as HM logic (see, e.g.,~\cite{BS:MMC}).

We start by introducing an infinitary version of the logic $\lcmu$,
which is then exploited to define fixpoint approximants.
Let $\lcmu^\infty$ denote an extension of $\lcmu$ with infinite
conjunctions, i.e., formulae of $\lcmu^\infty$ are defined by the
grammar
\begin{center}
    $
    \varphi\ ::=\  
    \p{X}{\vx} \  \mid\  
    \true\ \mid\ \bigwedge_{i \in I} \varphi_i\ \mid\ \neg\varphi\ \mid\ 
    \en{\vx}{\vy}{z}\, \varphi\  \mid\ 
    \ex{z}\varphi \mid\ \mu \p{X}{\vx}. \varphi\
    $
\end{center}
The semantics of $\lcmu^\infty$ is given as in
Definition~\ref{de:semantics-mu}, replacing the clause for conjunction
with $\sem[\mathcal{E}]{\bigwedge_{i \in I} \varphi_i}_\pi =
\bigcap_{i \in I} \sem[\mathcal{E}]{\varphi_i}_\pi \cap \ld{\bigwedge_{i \in I} \varphi_i}$. We denote by
$\lc^\infty$ the fragment of $\lcmu^\infty$ not including propositions
and fixpoint operators.

\begin{definition}[approximants]
  The $\alpha$-th approximant of a fixpoint formula in $\lcmu^\infty$, for an
  ordinal $\alpha$, is a formula in $\lc^\infty$, inductively defined
  as follows:
  \begin{quote}
    \begin{tabular}{lll}
      $\mu^0 \p{X}{\vx}. \varphi = \false$ \\[1mm]
      $\mu^{\alpha+1} \p{X}{\vx}. \varphi =  
      \varphi [\mu^\alpha \p{X}{\vx}. \varphi/X]$\\[1mm]
      $\mu^{\lambda} \p{X}{\vx}. \varphi =  
      \bigvee_{\alpha<\lambda} \mu^\alpha \p{X}{\vx}. \varphi$ & \qquad & for $\lambda$ a limit ordinal
    \end{tabular}
  \end{quote}
\end{definition}

A fixpoint formula $\mu \p{X}{\vx}. \varphi$ is intuitively
equivalent to the (infinite) disjunction of its approximants. More
formally:

\begin{lemma}[fixpoint unfolding via approximants]
  \label{le:approximants}
  Let $\mathcal{E}$ be a {\pes}. For any formula $\mu \p{X}{\vx}. \varphi$ in
  $\lcmu^\infty$ there exists an ordinal $\alpha$ such that 
  \begin{center}
    $\sem[\mathcal{E}]{\mu \p{X}{\vx}. \varphi}_\pi = \sem[\mathcal{E}]{\mu^\alpha \p{X}{\vx}. \varphi}_\pi$.
  \end{center}
\end{lemma}

\begin{proof}
  Recall that $\sem[\mathcal{E}]{\mu \p{X}{\vx}. \varphi}_\pi =
  \mathit{lfp}(f)$ where 
  $f : 2^{\ld{\p{X}{\vx}}} \to 2^{\ld{\p{X}{\vx}}}$
  is the function defined by $f(S) =
  \sem[\mathcal{E}]{\varphi}_{\pi[\p{X}{\vx} \mapsto S]}$.
  
  We already noted that the function $f$ is monotone in
  $2^{\ld{\p{X}{\vx}}}$ ordered by subset inclusion. Hence its least
  fixpoint can be obtained by iterating $f$ on $\emptyset$, the bottom
  element of the lattice, i.e., there exists an ordinal $\alpha$ such
  that $\mathit{lfp}(f) = f^{\alpha}(\emptyset)$, where
  $f^0(\emptyset) = \emptyset$, $f^{\alpha+1}(\emptyset) =
  f(f^\alpha(\emptyset))$ and $f^{\lambda}(\emptyset) =
  \bigcup_{\alpha < \lambda} f^\alpha(\emptyset)$ for $\lambda$ a limit
  ordinal.

  The observation that for any ordinal $\alpha$ it holds that
  $f^\alpha(\emptyset) = \sem[\mathcal{E}]{\mu^\alpha
    \p{X}{\vx}. \varphi}_\pi$ allows us to conclude. The latter can be
  proved by transfinite induction on $\alpha$.
  
  ($\alpha = 0$) 
  $\sem[\mathcal{E}]{\mu^0 \p{X}{\vx}. \varphi}_\pi =  
  \sem[\mathcal{E}]{\false}_\pi
  = \emptyset 
  = f^0(\emptyset)$

  \smallskip

  ($\alpha \to \alpha+1$) We have that
  \begin{quote}
    \begin{tabular}{lll}
      $\sem[\mathcal{E}]{\mu^{\alpha+1} \p{X}{\vx}. \varphi}_\pi =$ & \quad & [definition of $\mu^{\alpha+1} \p{X}{\vx}. \varphi$]\\[1mm]
      \quad $= \sem[\mathcal{E}]{\varphi [\mu^\alpha \p{X}{\vx}. \varphi/X]}_\pi =$ & & [Lemma~\ref{le:subst-mu}]\\[1mm]
      \quad $= \sem[\mathcal{E}]{\varphi}_{\pi[\p{X}{\vx} \mapsto \sem{\mu^\alpha \p{X}{\vx}. \varphi}_\pi]} =$ & \quad & [definition of $f$]\\[1mm]
      \quad $= f(\sem{\mu^\alpha \p{X}{\vx}. \varphi}_\pi) =$ & \quad & [inductive hypothesis] \\[1mm] 
      \quad $= f(f^\alpha(\emptyset))$
    \end{tabular}
  \end{quote}
  
  \smallskip
  
  ($\lambda$ limit ordinal) We have
  \begin{quote}
    \begin{tabular}{lll}
      $\sem[\mathcal{E}]{\mu^\lambda \p{X}{\vx}. \varphi}_\pi =$ & \quad &  [definition of $\mu^{\lambda} \p{X}{\vx}. \varphi$]\\[1mm]
      \quad $= \sem[\mathcal{E}]{\bigvee_{\alpha<\lambda} \mu^\alpha \p{X}{\vx}. \varphi}_\pi =$ & \quad &  [from Definition~\ref{de:semantics-mu}]\\[1mm]
      \quad $= (\bigcup_{\alpha<\lambda} \sem[\mathcal{E}]{\mu^\alpha \p{X}{\vx}. \varphi}_\pi) \cap \ld{\bigvee_{\alpha<\lambda} \mu^\alpha \p{X}{\vx}. \varphi} =$ & \quad &  [distributivity of $\cap$ w.r.t. $\cup$]\\[1mm]
      \quad $= \bigcup_{\alpha<\lambda} (\sem[\mathcal{E}]{\mu^\alpha \p{X}{\vx}. \varphi}_\pi \cap \ld{\bigvee_{\beta<\lambda} \mu^\alpha \p{X}{\vx}. \varphi})  =$ & \quad &  [$\ld{\bigvee_{\beta<\lambda} \mu^\alpha \p{X}{\vx}. \varphi} = \ld{\mu^\beta \p{X}{\vx}. \varphi}$ for \\
    & & any $\beta$, as all approximants have the same\\
    & & free variables]\\[1mm]
      \quad $= \bigcup_{\beta<\lambda} (\sem[\mathcal{E}]{\mu^\alpha \p{X}{\vx}. \varphi}_\pi \cap \ld{\mu^\alpha \p{X}{\vx}. \varphi}) =$ & \quad &  [since $\sem[\mathcal{E}]{\mu^\alpha \p{X}{\vx}. \varphi}_\pi \subseteq \ld{\mu^\alpha \p{X}{\vx}. \varphi}$]\\[1mm]
      \quad $= \bigcup_{\beta<\lambda} \sem[\mathcal{E}]{\mu^\alpha \p{X}{\vx}. \varphi}_\pi  =$ & \quad &  [by inductive hypothesis]\\[1mm]
      \quad $= \bigcup_{\alpha<\lambda} f^\alpha(\emptyset) =$ \\[1mm] 
      \quad $= f^\lambda(\emptyset)$
    \end{tabular}
  \end{quote}
  \qed

  



\end{proof}

We can finally prove that the logical equivalences induced by $\lc$
and $\lcmu$ are the same and they both coincide with $\hhpb$.

\begin{theorem}[invariance of logical equivalence]
  \label{th:coincidence}
  The logical equivalences of $\lc$ and $\lcmu$ coincide with $\hhpb$.
\end{theorem}

\begin{proof}
  First of all, since $\lcmu$ extends $\lc$, clearly $\equiv_{\lcmu}$
  implies $\equiv_{\lc}$ which in turn, by
  Proposition~\ref{pr:LogicToHhp}, implies $\hhpb$.  Hence
  $\equiv_{\lcmu}$ implies $\hhpb$.
  For the opposite direction, 
  note that Proposition~\ref{pr:HhpToLogic} can be
  straightforwardly adapted to logic $\lc^\infty$ (as finiteness of
  conjunction plays no role in the proof). Hence $\hhpb$ implies
  $\equiv_{\lc^\infty}$.
  An inductive argument, using Lemma~\ref{le:approximants}, allows one
  to show that for any closed formula in $\lcmu^\infty$ (and thus in
  particular any formula in $\lcmu$), there exists an equivalent
  formula in $\lc^\infty$, obtained by replacing all fixpoint
  operators with suitable approximants. Therefore
  $\equiv_{\lc^\infty}$ implies $\equiv_{\lcmu}$, hence  
  $\hhpb$ implies $\equiv_{\lcmu}$ as desired.
  \qed
\end{proof}

We conclude this section by mentioning that fragments of $\lcmu$
corresponding to fixpoint extension of step, pomset and history
preserving logic can be defined in the obvious way. The invariance of
logical equivalence for these fragments can be easily proved along the
lines of the previous proof.

\section{Conclusions: related and future work}
\label{se:conclude}

We have introduced a logic for true concurrency, which allows us to
predicate on events in computations and their mutual dependencies
(causality and concurrency). The logic subsumes standard HM logic and
provides a characterisation of the most widely known true concurrent
behavioural equivalences: hhp-bisimilarity is the logical equivalence
induced by the full logic, and suitable fragments are identified which
induce hp-bisimilarity, pomset and step bisimilarity.

As we mentioned in the introduction, there is a vast literature
relating logical and operational views of true concurrency, however,
to the best of our knowledge, a uniform logical counterpart of the true
concurrent spectrum was still missing. An exhaustive account of the
related literature is impossible; we just recall here the approaches
that most closely relate to our work.

In~\cite{DF:OLCM,LPS:TechRep94,Cher:BFbis} the causal structure of
concurrent systems is pushed into the logic.  The paper~\cite{DF:OLCM}
considers modalities which describe pomset transitions, thus providing
an immediate characterisation of pomset
bisimilarity. Moreover,~\cite{DF:OLCM,LPS:TechRep94,Cher:BFbis} show
that by tracing the history of states and adding the possibility of
reverting pomset transitions, one obtains an equivalence coarser than
hp-bisimilarity and incomparable with pomset bisimilarity, called weak
hp-bisimilarity.
Our logic intends to be more general by also capturing the interplay
between concurrency and branching, which is not observable at the
level of hp-bisimilarity.

The idea of studying logics for true concurrency, identifying suitable fragments which induce known or meaningful behavioural equivalences has been considered by several authors. 
In particular, a recent work~\cite{Gut:BMC} discusses a fixpoint modal
logic for true concurrent models, called separation fixpoint logics
(SFL), originally introduced in~\cite{Gut:LBGC}.  The logic SFL
includes modalities which specify the execution of an action causally
dependent/independent on the last executed one. Moreover, a
``separation operator'' deals with concurrently enabled actions.
%
%
%
This line of work is in turn inspired by the so-called
independence-friendly modal logic (IFML)~\cite{BF:IndepLogic02}, which
includes a modality that allows one to specify that the currently
executed action is independent from a number of previously executed
ones.
In this sense IFML is similar in spirit to our logic. 
Equivalences induced by (fragments of) IFML, with alternative
semantics, are investigated and shown to be often not standard in the
true concurrent spectrum. The fragment
of the logic in~\cite{Gut:BMC} without the separation operator
captures a weakening of hp-bisimilarity~\cite{Fro:pc}, which coincides
with hp-bisimilarity on a suitable subclass of safe Petri
nets~\cite{Gut:BMC}.
For similar reasons, the full logic induces an equivalence which is
weaker than hhp-bisimilarity, and incomparable with
hp-bisimilarity. Still a deeper comparison with this approach
represents an interesting open issue.

Several classical papers have considered temporal logics with
modalities corresponding to the ``retraction'' or ``backward''
execution of computations.  In
particular~\cite{JNW:BFOM,NC:GLNB,Ber:HHP,HS:PFP} study a so-called
path logic with a past tense (also called future perfect) modality:
the formula $@ \act\, \varphi$ is true when $\varphi$ holds in a state
which can reach the current one with an $\act$-transition. For systems
that do not exhibit autoconcurrency i.e., where events with the same
label are never enabled concurrently, such a logic can be shown to
characterise hhp-bisimilarity.
The restriction to systems without autoconcurrency can be relaxed by
modifying the past tense modality in a way which allows one to undo a
specific event executed in the past~\cite{NC:GLNB}. With such a
modification the logic becomes event-based logic, similar, in spirit
to our logic $\lc$.

Compared to these works, the main novelty of our approach resides in
the fact that the logic $\lc$ provides a characterisation of the different
standard true concurrent equivalences in a simple, unitary logical framework.
In order to enforce this view, we intend to pursue a formal comparison
with the logics for concurrency introduced in the literature. It is
easy to see that the execution modalities of \cite{Gut:BMC}
can be encoded in $\lc$ since they only refer to the last executed
event, while the formulae in $\lc$ can refer to any event executed in the
past. 
On the other hand, the ``separation operator''
of~\cite{Gut:BMC}, as well as the backward modalities
mentioned above (past tense, future perfect, reverting pomset
transitions) are not immediately encodable in $\lc$. A deeper
investigation would be of great help in shading further light on the
true concurrent spectrum. Moreover $\lc$ suggests an alternative,
forward-only, operational definition of hhp-bisimilarity which we
would expect to be closely related to the characterisation of
hhp-bisimilarity in~\cite{FH:PHHP}. This approach could be inspiring
also for other reverse bisimilarities~\cite{PhilUlid:REVERSE}.

Interestingly, the idea of considering a logic with event variables is
taken also in a very recent work~\cite{PU:LRM}, which provides an
elegant characterisation of (h)hp-bisimilarity via a logic, called
event identifier logic (EIL), with a backward execution
modality. The logic includes three operators: $\langle
x{:}\act\rangle\rangle$, $(x{:}\act)$ and $\langle \langle x
\rangle$. The formula $\langle x{:}\act\rangle\rangle \varphi$ holds
when, starting from the current configuration, an $\act$-labelled
event can be executed and, after the execution of such an event the
formula $\varphi$ holds. The formula $(x{:}\act) \varphi$ states that
the current configuration contains an $\act$-labelled event (which has
thus been executed in the past) and formula $\varphi$ holds. In both
cases, the $\act$-labelled event is bound to variable $x$ to be possibly
referenced in $\varphi$. Finally, $\langle \langle x \rangle$ holds
when the event bound to $x$ can be undone and then $\varphi$ holds.
The reason why both logics capture hhp-bisimilarity is conceptually
clear: the possibility of performing backward steps can be seen as a
mean of exploring alternative different futures. The very same
possibility is ``primitive'' in our logic where we can explore the
future of a configuration, without executing the corresponding events.
However, the formal relationships between EIL and our logic (e.g., the
possibility of encoding backward steps in our logic) is still to be
understood and represents a stimulating direction of future research.

As a byproduct of such an investigation, we foresee the identification
of interesting extensions of the concurrent spectrum, both at the
logical and at the operational side. For instance, it can be shown
that the fragment of $\lc$ where the operator $\en{\vx}{\vy}{z}$ is
restricted to bind $z$ to events consistent with those already
quantified induces an equivalence which admits a natural operational
definition, it is decidable and lies in between hp- and
hhp-bisimilarity, still being different from the equivalences
in~\cite{Gut:BMC}.

Connected to this, model-checking and decidability issues are
challenging directions of future investigation (see~\cite{Pen95} for a
survey of these issues over partial order temporal logics and logics
based on event structures having explicit operators representing
concurrency, causality and conflict). It is known that
hhp-bisimilarity is undecidable, even for finite state
systems~\cite{JNS:UDG}, while hp-bisimilarity is
decidable~\cite{Vog:DHPB,MP:MTS}. Characterising decidable fragments
of the logic could be helpful in drawing a clearer separation line
between decidability and undecidability of concurrent equivalences. A
promising direction is to impose a bound on the ``causal depth'' of
the future which the logic can quantify on. In this way one gets a
chain of equivalences, coarser than hhp-bisimilarity, which should be
closely related with $n$-hhp bisimilarities introduced and shown to be
decidable in~\cite{FH:PHHP}.
%
As for verification, we aim at investigating the
automata-theoretic counterpart of the logic. In previous papers,
hp-bisimilarity has been characterised in automata-theoretic terms
using HD-automata~\cite{MP:MTS} or Petri nets~\cite{Vog:DHPB}.
It seems that HD-automata~\cite{MP:MTS} could provide a suitable
automata counterpart of the fragment $\lchp$.
Also the game-theoretical approach proposed in~\cite{GB:MCG,Gut:BMC}
for the separation fixpoint logic as well as the model checking
techniques developed in~\cite{GW:MCPD} for their first order
$\mu$-calculus can be sources of inspiration.

Just note that the model checking problem is not trivial since it
may be the case that some formulae have infinite models only, even if
we limit ourselves to the finite fragment of the logic. For instance,
the formula $\enx[a]{}{}{w} \true \land\neg \en{}{}{x} \neg
\en{x}{}{y} \true$ only holds in an {\pes} which contains an infinite
causal chain of $\act$-labelled events.  Preliminary investigations
lead us to conjecture that model-checking is decidable on finite state
systems for the fixpoint extension of $\lchp$, $\lcp$ and $\lcs$.

\subsubsection*{Acknowledgements.}
We are grateful to Luca Aceto, Sibylle Fr\"oschle and to the anonymous
reviewers for their detailed comments and inspiring suggestions which
helped us in improving the the paper. In particular a remark from the
reviewers stimulated a more appropriate presentation of well-formed
formulae.

\bibliographystyle{alpha}
\bibliography{ConcLogic}


\newpage

\appendix

\section{Well-formed formulae}
\label{app:wf}

We identify a fragment of the logic $\lc$ where the restriction of
the denotations to include only legal pairs is enforced
syntactically. The idea is very simple: whenever we bind an event to a
variable we declare how it relates to all the events bound to the free
variables in the remaining part of the formula.

\begin{definition}[well-formed formulae]
  \label{de:well-formed}
  A formula $\varphi \in \lc$ is called \emph{well-formed} when, for
  any subformula of the kind $\en{\vx}{\vy}{z}\, \psi$,
  we have that $\fv{\psi} \subseteq \vx \cup \vy \cup
  \{z\}$. We denote by $\lcwf$ the fragment of $\lc$ consisting of
  well-formed formulae.
\end{definition}
Observe that any subformula of a well-formed formula is well-formed.

The semantics of well-formed formulae can be given as in
Definition~\ref{de:semantics}, without restricting to legal pairs. We
refer to this ``unrestricted'' semantics as the well-formed denotation of a formula.

\begin{definition}[semantics of well-formed formulae]
  \label{de:semantics-wf}
  Let $\mathcal{E}$ be a {\pes}. The well-formed denotation of a
  formula $\varphi$ in $\lcwf$, written $\sem[\mathcal{E}]{\varphi}_{\mathit{wf}}
  \in 2^{\conf{\mathcal{E}} \times \Env_{\mathcal{E}}}$ is defined inductively as
  follow:
  \begin{center}
    $
    \begin{array}{rcl}
      \sem[\mathcal{E}]{\true}_{\mathit{wf}} & = &  \conf{\mathcal{E}} \times
      \Env_{\mathcal{E}}
      \\[3mm]
      %
      %
      \sem[\mathcal{E}]{\varphi_1 \land \varphi_2}_{\mathit{wf}} & = &  
      \sem[\mathcal{E}]{\varphi_1}_{\mathit{wf}} \cap \sem[\mathcal{E}]{\varphi_2}_{\mathit{wf}}\\[3mm]
      \sem[\mathcal{E}]{\neg \varphi}_{\mathit{wf}} & = &  (\conf{\mathcal{E}} \times
      \Env_{\mathcal{E}}) \setminus \sem[\mathcal{E}]{\varphi}_{\mathit{wf}}\\[3mm]
      \sem[\mathcal{E}]{\en{\vx}{\vy}{z}\, \varphi}_{\mathit{wf}} & = &
      \{ (C, \eta) \mid 
      \begin{array}[t]{ll}
           \exists e \in \res{E}{C} \mathrm{\ such\ that }\\
           \lambda(e) = \act\ \land\ \eta(\vx) < e\ \land\ \eta(\vy) \conc e\\
           \land\  (C, \eta[z\mapsto e]) \in
           \sem[\mathcal{E}]{\varphi}_{\mathit{wf}}  & \}\\[3mm]
      \end{array}
      \\
      %
      %
      \sem[\mathcal{E}]{\ex{z}\, \varphi}_{\mathit{wf}} & = &
      \{ (C, \eta) \mid 
            \begin{array}[t]{ll}
          C \trans{\eta(z)} C'\ \land\  (C', \eta) \in
      \sem[\mathcal{E}]{\varphi}_{\mathit{wf}} & \}
      \end{array}

      \\[2mm]
      %
      %
    \end{array}
    $
  \end{center}
\end{definition}

The claim that the ``well-formedness'' is a syntactic counterpart of
the restriction to legal pairs is now formalised by proving that, for
closed well-formed formulae, the well-formed denotation given above and the
one based on legal pairs in Definition~\ref{de:semantics} do coincide.

\begin{proposition}[semantics of well-formed formulae]
  Let $\mathcal{E}$ be a {\pes}. Then, for any closed well-formed
  formula $\varphi$
  \[
  \sem[\mathcal{E}]{\varphi} = \sem[\mathcal{E}]{\varphi}_\mathit{wf}
  \]
\end{proposition}

\begin{proof}
  We can prove more generally that for any well-formed formula
  $\varphi$, it holds that
  \begin{center}
    $\sem[\mathcal{E}]{\varphi} =
    \sem[\mathcal{E}]{\varphi}_\mathit{wf} \cap \ld{\varphi}$.
  \end{center}
  From this the thesis immediately follows, since for a closed formula
  $\varphi$ it holds that $\ld{\varphi} = \conf{\mathcal{E}} \times
  \Env$. The proof can proceed by induction on $\varphi$.
  
  \medskip

  \noindent
  (case $\true$) 
  Since $\ld{\true} = \conf{\mathcal{E}} \times \Env$,
  we have 
  \begin{center}
    $\sem[\mathcal{E}]{\true}_\mathit{wf} \cap \ld{\true} =
    (\conf{\mathcal{E}} \times \Env) \cap
    (\conf{\mathcal{E}} \times \Env) = \conf{\mathcal{E}}
    \times \Env = \sem[\mathcal{E}]{\true}$.
  \end{center}
  
  \medskip

  \noindent
  (case $\varphi \land \psi)$ 
  We have
  \begin{quote}
    \begin{tabular}{ll}
      $\sem[\mathcal{E}]{\varphi \land \psi}_{\mathit{wf}} \cap \ld{\varphi \land \psi}$ & [by Definition~\ref{de:semantics-wf}]\\[1mm]
      \quad $= \sem[\mathcal{E}]{\varphi}_{\mathit{wf}} \cap
      \sem[\mathcal{E}]{\psi}_{\mathit{wf}} \cap \ld{\varphi \land \psi}$
      & [by inductive hypothesis]\\[1mm]
      \quad $=\sem[\mathcal{E}]{\varphi} \cap \ld{\varphi} \cap 
      \sem[\mathcal{E}]{\psi} \cap \ld{\psi}  \cap \ld{\varphi \land \psi}$ &
      [since $\ld{\varphi \land \psi} \subseteq \ld{\varphi} \cap \ld{\psi}$]\\[1mm]
      \quad $= \sem[\mathcal{E}]{\varphi} \cap \sem[\mathcal{E}]{\psi} \cap \ld{\varphi \land \psi}$ & [by Definition~\ref{de:semantics}]\\[1mm]
      \quad $= \sem[\mathcal{E}]{\varphi \land \psi}$ & 
    \end{tabular}
  \end{quote}
  \medskip

  \noindent
  (case $\neg \varphi$)
  We have
  \begin{quote}
    \begin{tabular}{ll}
      $\sem[\mathcal{E}]{\neg \varphi}_\mathit{wf} \cap \ld{\neg \varphi}$
      & [by Definition~\ref{de:semantics-wf}]\\[1mm]
      \quad $((\conf{\mathcal{E}} \times \Env) \setminus \sem[\mathcal{E}]{\varphi}_\mathit{wf}) \cap \ld{\neg \varphi}$
      & [since $\ld{\neg \varphi} = \ld{\varphi}$]\\[1mm]
      \quad $((\conf{\mathcal{E}} \times \Env) \setminus \sem[\mathcal{E}]{\varphi}_\mathit{wf}) \cap \ld{\varphi}$
      & [by calculation]\\[1mm]
      \quad $((\conf{\mathcal{E}} \times \Env) \cap \ld{\varphi}) \setminus (\sem[\mathcal{E}]{\varphi}_\mathit{wf} \cap \ld{\varphi})$
      & [by $\ld{\varphi} \subseteq \conf{\mathcal{E}} \times \Env$ and inductive hypothesis]\\[1mm]
      \quad $= \ld{\varphi} \setminus \sem[\mathcal{E}]{\varphi}$ &
      [by Definition~\ref{de:semantics}]\\[1mm]
      \quad $=  \sem[\mathcal{E}]{\neg \varphi}$ 
    \end{tabular}
  \end{quote}

  \medskip

  \noindent
  (case $\en{\vx}{\vy}{z}\, \varphi$)   
  By Definition~\ref{de:semantics-wf} we have
  \begin{quote}
      $\sem[\mathcal{E}]{\en{\vx}{\vy}{z}\, \varphi}_{\mathit{wf}} \cap \ld{\en{\vx}{\vy}{z}\, \varphi}=\\[1mm]
      \begin{array}{ll}
        = \{ (C, \eta) \mid & (C, \eta) \in \ld{\en{\vx}{\vy}{z}\, \varphi} \land\\[1mm] 
        & \exists e \in \res{E}{C}.\ 
       \lambda(e) = \act\ \land\ \eta(\vx) < e\ \land\ \eta(\vy) \conc e\
       \land\  (C, \eta[z\mapsto e]) \in
       \sem[\mathcal{E}]{\varphi}_{\mathit{wf}} \}
     \end{array}
     $
   \end{quote} 
   Now observe that, since the formula $\en{\vx}{\vy}{z}\, \varphi$ is
   well-formed, $\fv{\varphi} \subseteq \vx \cup \vy \cup \{z\}$ and
   thus $\fv{\en{\vx}{\vy}{z}\, \varphi} = \vx \cup \vy$. As a
   consequence, whenever $(C, \eta) \in \ld{\en{\vx}{\vy}{z}\,
     \varphi}$ and $e \in \res{E}{C}$ with $\eta(\vx) < e$ and
   $\eta(\vy) \conc e$, we have
   \begin{center}
     $e \cons \eta(\fv{\varphi} \setminus \{ z \})$ \quad and \quad $(C, \eta[z \mapsto e]) \in \ld{\varphi}$.
   \end{center}
   Therefore, we get
   \begin{quote}
     $\sem[\mathcal{E}]{\en{\vx}{\vy}{z}\, \varphi}_{\mathit{wf}} \cap \ld{\en{\vx}{\vy}{z}\, \varphi}=\\[1mm]
     \begin{array}{ll}
       = \{ (C, \eta) \mid & (C, \eta) \in \ld{\en{\vx}{\vy}{z}\, \varphi} \land\\[1mm] 
       & \begin{array}[t]{ll}
         \exists e \in \res{E}{C}.\ & e \cons \eta(\fv{\varphi} \setminus \{ z \})\ \land\ 
       \lambda(e) = \act\ \land\ \eta(\vx) < e\ \land\ \eta(\vy) \conc e\\[1mm]
       & \land\  (C, \eta[z\mapsto e]) \in
       \sem[\mathcal{E}]{\varphi}_{\mathit{wf}} \cap \ld{\varphi} \}
     \end{array}
   \end{array}
     $
   \end{quote} 
   Since by inductive hypothesis
   $\sem[\mathcal{E}]{\varphi}_{\mathit{wf}} \cap \ld{\varphi} =
   \sem[\mathcal{E}]{\varphi}$, we deduce that
   \begin{quote}
     $\sem[\mathcal{E}]{\en{\vx}{\vy}{z}\, \varphi}_{\mathit{wf}} \cap
     \ld{\en{\vx}{\vy}{z}\, \varphi}=
     \sem[\mathcal{E}]{\en{\vx}{\vy}{z}\, \varphi} $ 
   \end{quote}
   as desired.

   \medskip

  \noindent
  (case $\ex{z}\, \varphi$)
  We have
  \begin{quote}
    \begin{tabular}{ll}
      $\sem[\mathcal{E}]{\ex{z}\, \varphi}_\mathit{wf} \cap \ld{\ex{z}\, \varphi}$
      & [by Definition~\ref{de:semantics-wf}]\\[1mm]
      \quad $= \{ (C, \eta) \mid 
                 C \trans{\eta(z)} C'\ \land\  (C', \eta) \in
                 \sem[\mathcal{E}]{\varphi}_{\mathit{wf}}  \} \cap \ld{\ex{z}\, \varphi}$
      & [by calculation]\\[1mm]
      \quad = $\{ (C, \eta) \mid (C, \eta) \in \ld{\ex{z}\, \varphi}\ \land\ 
                 C \trans{\eta(z)} C'\ \land\  (C', \eta) \in
                 \sem[\mathcal{E}]{\varphi}_{\mathit{wf}}  \}$\\
      \multicolumn{2}{r}{[since $(C, \eta) \in \ld{\ex{z}\, \varphi}\ \land\ 
      C \trans{\eta(z)} C'$ iff
      $(C', \eta) \in \ld{\varphi}\ \land\ C \trans{\eta(z)} C'$]}\\[1mm]
      \quad $= \{ (C, \eta) \mid 
                 C \trans{\eta(z)} C'\ \land\  (C', \eta) \in
                 \sem[\mathcal{E}]{\varphi}_{\mathit{wf}} \cap \ld{\varphi} \} $
      & [by inductive hypothesis]\\[1mm]
      \quad $= \{ (C, \eta) \mid 
                 C \trans{\eta(z)} C'\ \land\  (C', \eta) \in
                 \sem[\mathcal{E}]{\varphi} \} $
      & [by Definition~\ref{de:semantics-wf}]\\[1mm]
      \quad $= \sem[\mathcal{E}]{\ex{z}\, \varphi}$
    \end{tabular}
  \end{quote}  
  \qed
\end{proof}

Restricting to well-formed formulae does not alter the logical equivalence which remains hhp-bisimilarity.

\begin{proposition}[well-formed formulae induce hhp-bisimilarity]
   Let $\mathcal{E}_1$ and $\mathcal{E}_2$ be {\pes}s. Then
  $\mathcal{E}_1 \hhpb \mathcal{E}_2$ iff $\mathcal{E}_1 \equiv_{\lcwf}
  \mathcal{E}_2$.
\end{proposition}
 
\begin{proof}
  The fact that if $\mathcal{E}_1 \hhpb \mathcal{E}_2$ then
  $\mathcal{E}_1 \equiv_{\lcwf} \mathcal{E}_2$ follows immediately by
  Proposition~\ref{pr:HhpToLogic}, since $\lcwf$ is a fragment of
  $\lc$.

  The converse implication can be proved essentially as for the full
  logic $\lc$ (Proposition~\ref{pr:LogicToHhp}) since the restriction to
  well-formed formulae smoothly integrates in the proof. More in
  detail, most of the proof of Proposition~\ref{pr:LogicToHhp},
  remains unchanged. When showing that relation $R$ is an
  hp-bisimilarity, it is sufficient to note that if the formulae
  $\psi^i$ are assumed to be well-formed then also the newly
  constructed formula $\varphi = \en{\vx}{\vy}{x_e}(\ex{X_{C_1}}
  \ex{x_e} \true \land \psi^1\land\ldots\land\psi^n)$ is
  well-formed. In fact, by construction $\vx, \vy \subseteq X_{C_1}$
  are such that $\eta_1(\vx)$ is the set of causes of $e$ in $C_1$ and
  $\eta_1(\vy)$ is the set of events in $C_1$, hence $\vx \cup \vy =
  X_{C_1}$.
  Moreover $\fv{\psi^i}\subseteq X_{C_1'}=X_{C_1}\cup\{x_e\}$ and thus
  $\fv{\ex{X_{C_1}} \ex{x_e} \true \land \psi^1\land\ldots\land\psi^n}
  = X_{C_1} \cup \{ x_e \}$. Hence $\varphi$ is well-formed.
  \qed
\end{proof}

The entire theory, including the fragments for step, pomset and
hp-bisimilarity and the logic with recursion could be developed
alternatively by focusing on the well-formed fragment of the logic,
with the well-formed semantics.

\end{document}